\documentclass[a4paper,11pt]{article}
\pdfoutput=1

\usepackage{jheppub}

\usepackage{xspace}
\usepackage{pslatex}
\usepackage{amsmath}
\usepackage{txfonts}

\usepackage{cancel}

\graphicspath{{./}}

\def\mt{m_t}

\def\nn{\nonumber}

\def\min{{\rm min}}
\def\true{{\rm true}}

\def\NLO{{\textrm{NLO}}}
\def\Refs#1{refs.~\cite{#1}}
\def\Ref#1{ref.~\cite{#1}}
\def\Eq#1{eq.~(\ref{#1})}
\def\Fig#1{figure~\ref{#1}}
\def\Lboost#1{\ensuremath{\Lambda^b(#1)}}
\def\Lrotx#1{\ensuremath{\Lambda_x^r(#1)}}
\def\Lroty#1{\ensuremath{\Lambda_y^r(#1)}}

\def\Kallen#1{\lambda\left(#1\right)}
\def\ycut{y_{\mbox{\scriptsize cut}}}
\def\yij{{y_{ij}}}
\def\ycut{\ensuremath{y_{\textrm{cut}}}}
\def\MEM{Matrix Element Method\xspace}

\title{Extending the \MEM beyond the 
  Born approximation: Calculating event weights at 
  next-to-leading order accuracy.}
 
 \author{Till Martini}
 \author{and Peter Uwer}
 \affiliation{Humboldt-Universit\"at zu Berlin, Institut f\"ur Physik,\\
  Newtonstra{\ss}e~15, 12489~Berlin, Germany}
  
\emailAdd{Till.Martini@physik.hu-berlin.de}  
\emailAdd{Peter.Uwer@physik.hu-berlin.de}


\arxivnumber{1506.08798}

\subheader{\hfill HU-EP-15/28,  JHEP09(2015)083}

\abstract{
  In this article we illustrate how event weights for jet events can
  be calculated efficiently at next-to-leading order (NLO) accuracy in
  QCD.
  This is a crucial prerequisite for the application of the Matrix Element Method in
  NLO. We modify the recombination procedure used in jet algorithms,
  to allow a factorisation of the phase space for the real corrections
  into resolved and unresolved regions. Using an appropriate infrared
  regulator the latter can be integrated numerically. As illustration,
  we reproduce differential distributions at
  NLO for two sample processes. As further application and proof of concept, we apply the Matrix Element Method
  in NLO accuracy to the mass determination of top quarks produced in
  $e^+e^-$ annihilation. This analysis is relevant for a future Linear
  Collider. We observe a significant shift in the extracted mass
  depending on whether the Matrix Element Method is used in leading or next-to-leading
  order.
}

\begin{document}
\maketitle

\section{Introduction}
With the steadily increasing computing power multivariate methods are
nowadays standard techniques in the experimental analysis. Initially
introduced in experimental studies where the event rates are
small and signals are difficult to disentangle from overwhelming
backgrounds, multivariate methods have been proven useful also in a
more general context. Among these methods the \MEM represents a
prominent example, since it allows a direct comparison of observed
event samples with expectations within a specific theoretical model.
As originally introduced in \Refs{Kondo:1988yd,Kondo:1991dw} the
method is based on the assumption that the probability to observe a
specific event can be calculated using the corresponding matrix
element together with so-called transfer functions which model the
probability for the observation of a given partonic event as a
specific (hadronic) event at the detector level. It has been argued
that if all the ingredients in this procedure are known with optimal
accuracy, the event likelihood defined in this way represents an
optimal statistical test. The method thus makes maximal use of the
information contained in the single event.  Based on this assumption
the \MEM has been used for example to measure the top-quark mass at
the Tevatron (see for example
\Refs{Abbott:1998dn,Abazov:2004cs,Abulencia:2006mi}).  Alternatively
the \MEM can be used to distinguish different
hypotheses like for example background versus signal hypothesis or SM
versus BSM physics hypothesis for a given event sample (see for example 
\Ref{Artoisenet:2013vfa,Khiem:2015ofa}). Let us also
mention that the \MEM exists today in different flavors like for
example the MELA approach as used for example 
in \Refs{Gao:2010qx, Bolognesi:2012mm}. The construction of optimal
observables as used for example in 
\Refs{Diehl:1993br,Diehl:1996wm,Janot:2015yza} may also be seen as a
variation of the same theme.
So far in
most cases the method is only applied using leading-order (LO) matrix
elements. To simplify the application of the \MEM the automated
calculation of the required event weights has been studied recently in
\Ref{Artoisenet:2010cn}.  Given the recent progress concerning the
calculation of NLO corrections it is natural to include also NLO
corrections in the evaluation of the matrix elements. A first attempt
in this direction has been made in \Ref{Alwall:2010cq} where the
effect of QCD radiation has been studied.
In \Refs{Soper:2011cr,Soper:2012pb} the radiation pattern of boosted Higgs bosons and top quarks is studied and compared with the radiation profile of QCD jets. 
In \Ref{Soper:2014rya} the information from the hard matrix element and a parton shower is used  for a signal versus background discrimination for the signal process $Z'$ decaying to boosted top quarks. 
In \Refs{Campbell:2012cz,Campbell:2012ct,Campbell:2013uha} the impact of NLO
corrections including also the virtual corrections is investigated and a
possible extension of the \MEM beyond the Born approximation is
discussed. A first detailed application has been presented in
\Ref{Campbell:2013hz} where Higgs production with subsequent decay
into $H\to Z\gamma$ is investigated. So far the method presented in
\Refs{Campbell:2012cz,Campbell:2012ct} is restricted
to the production of uncolored objects and the extension to the
production of colored particles like for example top-quark pairs is
still missing. In \Ref{Campbell:2013uha} the extension 
to include hadronic production of jets is investigated by means
of a longitudinal boost along the beam axis to remove the
unbalanced transverse momentum and map NLO and LO jets.

The extension of the final state phase space encountered in the
generalisation of the \MEM beyond the Born approximation is an
intrinsic problem which makes the NLO extension non-trivial. Real
corrections which appear as additional contribution, when higher order
corrections are taken into account, allow for the emission of an
additional parton.  In addition to the $2 \to n$ Born kinematics also
contributions living on an $n+1$ parton phase space need thus to be
considered.  Restricting the attention to jet physics, one may argue
that only regions of the extended phase space in which the additional
parton is clustered/merged into a jet contribute to $n$-jet
observables and we end up again with a $2 \to n$ configuration---now
in terms of jets instead of partons.  However, applying standard jet
algorithms, the resulting jets typically do not satisfy the
kinematical constraints of the Born process. In particular, the
clustering will in general create a non vanishing mass for the jets
even for massless partons. Also momentum conservation is not
necessarily required by all jet algorithms. It is thus not clear how
the contribution from the real corrections can be unambiguously
combined with the virtual corrections which respect the Born
kinematics.  Furthermore, the practical calculation of next-to-leading
order event weights---which must take into account virtual as well as
real corrections---is non-trivial. As we shall describe in more detail
in the next section, the problem is related to the phase space
integration of real corrections which lead to an $n$-particle final
state after clustering. Defining event weights for `jet-events' in
NLO accuracy thus requires two aspects to be addressed:
\begin{enumerate}
\item In the theoretical predictions a modified clustering may be used
  which guarantees momentum conservation and keeps the clustered jets
  on-shell. More precisely, the mass of a jet which is formed through
  a merging of a massless parton with another parton should be equal
  to the mass of the radiating parton.  Merging two massless partons
  should result in a massless jet. It is worth stressing that also for
  established jet algorithms, which rely on a $2\to 1$ clustering, the
  jet masses created in the perturbative calculation have very little
  to do (in case of massless partons) with the jet masses observed in
  the experimentally measured jets in case of `light' 
  jets.\footnote{This statement does not apply to fat jets or highly
  boosted objects.} 
  The latter are mostly
  related to non-perturbative effects. In that sense, a modified
  clustering as proposed here is equally well motivated as what is
  currently used. In fact, one may even argue that a clustering
  fulfilling the above constraints may lead to a better separation of
  perturbative and non-perturbative physics.

\item A method needs to be constructed allowing the efficient
  integration of the regions in the $n+1$-parton phase space contributing
  to the considered $n$-jet configuration. (In principle an inclusive 
  observable may also receive contributions from $n+1$ jet configurations
  requiring the evaluation of the corresponding weights. In this case
  the problem is however very similar to the leading-order situation.) 
\end{enumerate}
In this article we illustrate that both aspects can be addressed by
using an appropriate jet clustering which is intimately related to a
factorised parameterisation of the $n+1$-parton phase space into an
$n$-particle phase space of the $n$ jets times the phase space related
to the `clustered' (unobserved) parton. The general idea is to extend
the typical $2\to 1$ clustering to a $3\to 2 $ clustering as it is
well known for example from the dipole subtraction method
\cite{Catani:1996vz,Catani:2002hc}.  This clustering satisfies
momentum conservation as well as on-shell conditions at the expense of
introducing an additional spectator which allows to guarantee momentum
conservation which would be otherwise violated by enforcing the
on-shell condition. Some freedom exists how to choose the additional
`spectator'. For example, to minimise the difference to the
traditional clustering, one may choose the spectator such that the
momentum reshuffling is minimised. Having chosen the spectator, a
recombination according to the Catani-Seymour phase space
factorisation \cite{Catani:1996vz,Catani:2002hc} is applied. In this
way, the reduced kinematics appearing in the Catani-Seymour
subtraction formalism can be identified with the final state jets. At
the same time, the factorised phase space can be integrated over the
unresolved regions to obtain the contributions from the real
corrections to the event weight. We will give more details in the next
section.

It should be noted that using the four mappings given in the
Catani-Seymour subtraction algorithm as clustering prescriptions in
the proposed $3\rightarrow 2$ jet algorithm is sufficient, to construct
appropriate jet algorithms covering most of the relevant collider
experiments: production of at least 2 jets at a lepton collider, deep
inelastic scattering and hadronic production of electroweak final
states and/or jets.  The respective final states can have arbitrary
masses.

Let us briefly compare the method outlined above with some existing
work where similar ideas have been applied in a different context.
For example, the ideas presented here share some features with
\Ref{Weinzierl:2001ny}. There are however important differences. In
\Ref{Weinzierl:2001ny} an additional resolution parameter is
introduced to define `resolved' partons similar to what is done in the
phase space slicing method \cite{Giele:1991vf,Giele:1993dj}. Using the
resolved partons in an intermediate step any physical jet algorithm
should in principle be applicable. Care has to be taken that the two
cuts---the artificial one to define resolved partons and the physical
one to define jets within a given jet algorithm---do not interfere. In
the approach described above we work directly with jets defined 
by the jet algorithm used in the experimental
analysis. In \Ref{Weinzierl:2001ny} only final state singularities in
$e^+e^-$ annihilation are considered. Here we include initial state
singularities as well. Technically, only one phase space mapping is
required in \Ref{Weinzierl:2001ny}. As we will see in the next
section, in general more mappings are required and the application of
the Catani-Seymour subtraction method is less obvious.  In
\Refs{Dinsdale:2007mf,Schumann:2007mg} the implementation of a parton
shower based on the Catani-Seymour subtraction method is studied.
Although the aim of this work is rather different, the
parameterisation of the phase space is similar to what is used in this
article and many useful results which we collect in the following can
be found also in \Refs{Dinsdale:2007mf,Schumann:2007mg}.\\
In \Ref{Giele:2011tm} a method to generate the phase space of $n+1$ massless 
particles  by forward branching of configurations of $n$ massless particles is presented. 
This method which is applied in 
\Refs{Campbell:2012cz,Campbell:2012ct,Campbell:2013uha,Campbell:2013hz}
employs two $3\rightarrow 2$ prescriptions to cluster three massless 
partons to two massless jets. The method that we present in this article can
be seen as a generalisation of this approach with two further
prescriptions for the clustering and the extension to massive particles.
As a proof of concept, especially for the aforementioned generalisations,
we apply the Matrix Element Method in NLO accuracy to a process with massive
colored particles in the final state. 

The article is organised as follows. In section
\ref{sec:PhaseSpaceparameterisation} we present the phase space
parameterisation used in the numerical integration. Roughly speaking
the Catani-Seymour mapping is inverted.  In section \ref{sec:checks}
we validate the approach by various cross checks.  As a proof of
concept we illustrate in section \ref{sec:application} the \MEM
including NLO corrections applied to top-quark pair production in
$e^+e^-$ annihilation. We summarise our main findings in the
conclusion in section \ref{sec:conclusion}.

\section{Formalism}
\label{sec:formalism}

\subsection{\MEM and event weights at next-to-leading order} 

The \MEM tries to make maximal use of the information
provided by an individual event. Instead of considering distributions,
calculated for event samples, the probability of the event in the
context of a given theory is investigated. In what follows we assume that 
all experimentally available information of the event is collected
in the variable $\vec{x}$. In the ideal situation that all momenta have been
reconstructed, one may think of $\vec{x}$ as the collection of the
observed momenta which we label with $J_1,\ldots,J_n$:
\begin{equation}
 \vec{x}=(J_1,\ldots,J_n). 
\end{equation}
However, since some particles may escape detection or are only partially
reconstructed, the experimentally accessible information may be in practice
only a subset of this information. 

Putting aside for the moment higher order corrections,
one may interpret the partonic cross section calculated for a
specific model---for example the Standard Model---as the probability
distribution to observe a partonic event.     
A model-dependent likelihood, with model parameters $\vec{\Omega}$, for
observing an event $\vec{x}$ is than given schematically by
\begin{equation}\label{eq:likelihood}
 {\cal L}(\vec{x}\;|\,\vec{\Omega}) = 
 \frac{1}{\sigma}\int\!\! dy_1\ldots dy_n\;\frac{d^n\sigma}{dy_1\ldots
 dy_n}
 \times W(\vec{x},\vec{y}),
\end{equation}
where the differential cross section is denoted by $d^n\sigma / dy_1\ldots
 dy_n$ and
the so-called transfer function $W(\vec{x},\vec{y})$ describes the probability
that a partonic event $\vec{y}$ is measured in the detector as the event
$\vec{x}$. In principle, the variables collected in $\vec{y}$ may be
chosen independently from the variables in $\vec{x}$. Even the
 dimension of the two vectors does not need to agree. However, it may
prove beneficial to chose the two sets as closely related as possible.
Assuming that the two can be identified, an ideal detector would than
correspond to the situation in which the transfer function is 
given by a delta function: $W(\vec{x},\vec{y})=\delta(\vec{x}-\vec{y})$.

Roughly speaking, maximising the likelihood with respect to
$\vec{\Omega}$ for a given event sample gives an estimator for the
model parameter.  This is the essence of the so-called \MEM (MEM).
(More details can be found in
\Refs{Kondo:1988yd,Kondo:1991dw,PhysRevD.45.1531,Abbott:1998dn,%
  EstradaVigil:2001eq,Abulencia:2006mi,Abazov:2008kt,Fiedler:2010sg,%
  Artoisenet:2010cn,Gainer:2013iya,Aad:2015gra,Khiem:2015ofa}.) 
Since all the information available in the single
measurement is retained, this approach is believed to make maximal use
of the information content of the single event.  
  
While in principle the integration over the transfer
functions looks straightforward, in practice it is not trivial due to
the peak structure of the transfer functions. In addition, we note
that the transfer functions need to be determined within the
experimental analysis, which may also represent a non-trivial task. In
the following we do not consider this issue any further since current
experimental analyses using the \MEM are already used
to this type of problem. The focus of this article is the extension of
the MEM beyond the Born approximation. To be specific we assume
$\vec{y}$ in the following to be the collection of final state
momenta---which may be obtained in the case of the real corrections
through the merging of collinear or soft partons according to a
specific clustering procedure.  To distinguish these momenta from the
partonic momenta we call them jet-momenta in what follows. They may be
seen as the perturbative approximation of the jets observed in the
experiments. (More
details will be given in the next subsection.) The motivation to use
the jet-momenta is threefold:
\begin{enumerate}
\item Being differential in the jet-momenta all relevant information about the
  differential cross section is kept, allowing in a second step also to
  use a different set of variables.
\item Using the jet momenta, identifying the transfer functions with
  delta-functions may provide a reasonable first approximation.
\item Being able to calculate event weights for `jet events' including
  higher order corrections is interesting on its own right.
\end{enumerate}
In the rest of this subsection we illustrate the main obstacle in the
calculation of event weights for jet events when higher order
corrections are included.  To start we consider first the cross
section in Born approximation.  Although not useful in practice, we
may write the differential cross section in terms of jet-momenta by
introducing delta-functions:
\begin{eqnarray}
  {d\sigma}\over{d^4J_1\dots d^4J_n} &=& {1\over 2 s} 
  \int dR_n(p_a+p_b,p_1,\ldots,p_n) |{\cal
    M}(p_a,p_b,p_1,\ldots,p_n)|^2 \nonumber\\
  & &\times\delta(J_1 - \tilde J_1(p_1,\ldots,p_n)) \ldots 
  \delta(J_n - \tilde J_n(p_1,\ldots,p_n)),
\end{eqnarray}
where $s$ denotes the center of mass energy squared, $|{\cal M}|^2$ is
the squared matrix element and $dR_n$ is the Lorentz invariant phase
space measure
\begin{equation}
  \label{eq:phasespace}
   dR_{n}(P,p_1,\ldots,p_n) = (2\pi)^4\delta(P - \sum_i p_i)
   \prod_{i=1}^n {d^4p_i\over (2\pi)^3} \delta_+(p_i^2-m_i^2),
\end{equation}
where $m_i$ denotes the mass of the $i$-th parton. The incoming
particles with momenta $p_a$ and $p_b$ are assumed colorless. In case
strongly interacting particles are considered in the initial state 
additional convolutions with the parton distribution functions need to
be introduced. The momenta of the final state partons are given by
$p_1,\ldots,p_n$. The functions $\tilde J_i(p_1,\ldots,p_n)$ describe
how the jet momenta are calculated from the parton momenta. Since in
leading-order no recombination is possible, the jet momenta are
identified with the parton momenta:
\begin{equation}
  \tilde J_i(p_1,\ldots,p_n) = p_i.
\end{equation}
Obviously, it is than straightforward to evaluate the delta-functions and
obtain the differential cross section in terms of the jet-momenta.
Including next-to-leading order corrections, we need to consider the
contribution from virtual corrections as well as real
corrections. Ignoring for the moment the fact that both contributions
are individually divergent due to soft and collinear singularities, we may apply
the same argument as above to calculate the virtual corrections to
the differential cross section. For the real
corrections, however, the situation becomes more complicated. If we
ask for precisely $n$ jets in the final state, we need to integrate over the
regions of the  $n+1$ parton phase space in which $n+1$ partons are clustered
to $n$ jets. More precisely we need to evaluate integrals of the form
\begin{eqnarray}
   &\int & dR_{n+1}(p_a+p_b,p_1,\ldots,p_{n+1}) |{\cal
    M}(p_a,p_b,p_1,\ldots,p_{n+1})|^2 \nonumber\\
  & &\times\delta(J_1 - \tilde J_1(p_1,\ldots,p_{n+1})) \ldots 
  \delta(J_n - \tilde J_n(p_1,\ldots,p_{n+1}))
    \Theta_{\mbox{\scriptsize n-jet}}
\end{eqnarray}
where the functions $\tilde J_n$ encode now how the $n+1$ partons are
clustered to $n$ jets and $\Theta_{\mbox{\scriptsize n-jet}}$
restricts the integration to the $n$ jet region. (We note that the
inclusion of the $n+1$ jet region is straightforward since no
clustering occurs.)
The functions $\tilde
J_n$ depend on the phase space region through the recombination
procedure since in different phase space regions different partons
are merged to form a jet. Evidently, the delta functions cannot be
integrated numerically and an analytic approach is required. This is
one facet of the problem we address in the next subsection. There is,
however, a further problem: Using the standard recombination
procedure, which is often simply the sum of the four momenta of the
merged objects, we obtain in general massive jets even in case that we
started with massless objects. As mentioned in the introduction it is
thus a priori not clear how the contribution of the real corrections
can be combined point-wise with the virtual corrections where the jets
may have different masses. This is, however, required to define an
event-weight with NLO accuracy. In the next section we will show how
the two issues are connected and can be addressed by a modification of
the clustering prescription.  In particular, we show how---by using a
modified recombination procedure---the `real' phase space can be
factorised into an $n$ jet phase space and a remainder with the
property that the $n$ jet phase space preserves the Born kinematics.
As long as the transfer function is not approximated by a
$\delta$-function one could in principle relax the requirement to map
the unresolved regions of the real corrections onto the Born
kinematics,
since \Eq{eq:likelihood} may be calculated for an arbitrary set of
partonic variables used to describe virtual or real
corrections. However, using the aforementioned identification of the
phase space for real and virtual corrections allows 
to define point wise event weights in NLO accuracy.
It is then straightforward to generalise the \MEM\ to NLO: The set of $\vec{y}$ variables in 
\Eq{eq:likelihood} are just reinterpreted as describing `theory jets'
as introduced before.   No further extension of \Eq{eq:likelihood} is required.
Note that this approach is meant to be applied in fixed order: Parton
shower corrections which partially resum higher order corrections
would lead to a double counting if naively included in this approach.

\subsection{A modified recombination prescription and 
phase space factorisation }
Using the resolution $\yij$ to define the jet
function $F^n_{J_1,\ldots,J_n}(p_1,\ldots,p_{n+1})$ 
we may write for the $n$-parton final state
\begin{equation}
  F^n_{J_1,\ldots,J_n}(p_1,\ldots,p_{n}) = \prod_{i\not =j} 
  \Theta(\yij(p_1,\ldots,p_n) -\ycut).
\end{equation}
The resolution depends on the final state objects: partons in
case of the perturbative calculation and hadrons in case of the
experimental analysis.  $\ycut$ defines a preset value for the jet
resolution. The momenta $p_1,\ldots,p_n$ refer to the parton momenta,
while $J_1,\ldots,J_n$ refer to the jet momenta. We have included them
in the definition of the jet function since every jet algorithm
includes in addition to the resolution also a prescription
how to define the momenta of a jet in case it is formed by the merging
of two finale state objects. In leading-order the jet momenta are
identified with the parton momenta, since no merging is possible.  In
NLO we also need the jet function for the case that $n+1$ partons form
$n$ jets.  Since for soft and collinear configurations the jet
function $F^{n+1}_{J_1,\ldots,J_n}$ needs to reproduce
$F^n_{J_1,\ldots,J_n}$ to ensure the cancellation of soft and
collinear divergencies, the jet function may be written as
\begin{equation}
\label{eq:JetAlg-1}
  F^{n+1}_{J_1,\ldots,J_n}(p_1,\ldots,p_{n+1}) = \sum_{i\not= j}
  \Theta(\ycut-\yij(p_1,\ldots,p_{n+1})) 
  F^{n}_{J_1,\ldots,J_n}(J_1,\ldots,J_n)
\end{equation}
with
\begin{equation}
  \Theta(\ycut-\yij(p_1,\ldots,p_{n+1})) = 1
\end{equation}
for soft or collinear configurations. (For the moment we ignore initial
state singularities. As we shall see the extension to include them as
well is straightforward.) The step functions assure that a
recombination of two partons into one jet occurs. For each possible
combination $i,j$ with $\yij<\ycut$ the momenta $J_i$ are obtained
through the respective recombination procedure from the original
momenta $p_i$. As mentioned before the mapping should respect momentum
conservation and keep the recombined particle on the respective {\it
  mass-shell} in the sense defined above. This can be achieved by
using the mapping introduced in the Catani-Seymour subtraction method.
Depending on the unresolved partons and the chosen spectator four
different mappings are given in \Ref{Catani:1996vz,Catani:2002hc}.
For each unresolved configuration we may choose for example the
combination with the smallest momentum transfer to the spectator
parton. More general we may define functions
\begin{equation}
  \Theta_{k}(p_1,\ldots,p_{n+1}),\, k=1,\ldots,n+1
\end{equation}
with the requirement that
\begin{equation}
  \sum_{k} \Theta_{k} = 1
\end{equation}
to select a specific mapping. For the numerical phase space
integration using Monte Carlo methods it might be useful to use smooth
functions $\Theta_k$ instead of step functions. 

Introducing $\sum_{k} \Theta_{k} = 1$ in \Eq{eq:JetAlg-1} we get
\begin{equation}
    F^{n+1}_{J_1,\ldots,J_n}(p_1,\ldots,p_{n+1}) = \sum_{i\not = j,
  k\not=i,j}
  \Theta(\ycut-\yij(p_1,\ldots,p_{n+1})) \Theta_k
  F^{n}_{J_1,\ldots,J_n}(J_1,\ldots,J_n)
\end{equation}
For the jet cross section the contribution from the real 
corrections reads
\begin{eqnarray}
  d\sigma &=& {1\over 2s} \sum_{i\not = j,
  k\not=i,j}
  \Theta(\ycut-\yij(p_1,\ldots,p_{n+1})) \Theta_k
  F^{n}_{J_1,\ldots,J_n}(J_1,\ldots,J_n) \nn\\
  & &\times |{\cal M}_{n+1}|^2 
  dR_{n+1}(p_1,\ldots,p_{n+1}),
\end{eqnarray}
with the phase space measure as defined in \Eq{eq:phasespace}.
 In what follows we
consider the functions $\Theta_k$ as part of the jet algorithm. The
role of the $\Theta_k$ is to select in each phase space region where 
partons/jets are merged the appropriate clustering.
Since in each region $\Theta_k$ selects a mapping $(p_1,\ldots,p_{n+1})
\to (J_1,\ldots,J_n)$ we may change the integration variables
accordingly using the respective Catani-Seymour parameterisation of the
phase space:
\begin{eqnarray}
  d\sigma &=& {1\over 2s} \sum_{i\not = j,
  k\not=i,j}
  \Theta(\ycut-\yij(p_1,\ldots,p_{n+1})) \Theta_k
  F^{n}_{J_1,\ldots,J_n}(J_1,\ldots,J_n) \nn\\
  & &\times |{\cal M}_{n+1}|^2 
  dR_{n}(J_1,\ldots,J_n)dR_{ij,k}
\end{eqnarray}
where $dR_{ij,k}$ denotes the respective phase space measure
introduced in \Ref{Catani:1996vz,Catani:2002hc}.  Note that
$p_1,\ldots,p_{n+1}$ appearing in the jet function and the matrix
elements should be expressed in terms of the the momenta
$J_1,\ldots,J_n$ and the integration variables used in $dR_{ij,k}$.
Using the factorised phase space it is straightforward
to calculate the contribution of the real corrections to
the event weight in NLO accuracy:
\begin{equation}
  \label{eq:MasterEq}
  {1\over 2s} \sum_{i\not = j,
  k\not=i,j}
  \Theta(\ycut-\yij(p_1,\ldots,p_{n+1})) \Theta_k(p_1,\ldots,p_{n+1})
   |{\cal M}_{n+1}|^2 
  dR_{ij,k}.
\end{equation}
In the integration the momenta $p_1,\ldots,p_{n+1}$ are determined
from the jet momenta $J_1,\ldots,J_n$ and the variables used in
$dR_{ij,k}$. The inversion of the mapping $(p_1,\ldots,p_{n+1}) \to
(J_1\ldots,J_n, \Phi)$ where $\Phi$ denotes the collection of
variables used in the (unresolved) phase space measure $dR_{ij,k}$ is
discussed in section~\ref{sec:PhaseSpaceparameterisation}.

We have ignored so far that the phase space integration is in general divergent
due to soft and collinear singularities. Since the
Catani-Seymour phase space factorisation is valid in $d$ dimension it
is straightforward to regularise the divergencies within dimensional
regularisation. Conceptually the easiest way to deal with the
singularities is to apply a phase space slicing
\cite{Giele:1991vf,Giele:1993dj}. In the numerical integration the
integration over the unresolved parton is cut-off to avoid the
collinear and soft configurations. In the singular regions, soft and
collinear factorisation can be used to simplify the matrix elements
such that the integration can be done analytically. The singularities
obtained in this way are then combined with the virtual corrections.

As far as the application of the Catani-Seymour subtraction method is
concerned the situation is more involved: To
allow the combination of the (integrated) subtraction term with the
virtual corrections the jet algorithm or in general the observable is
evaluated for the reduced kinematics in the Catani-Seymour formalism.
The term which is added (and subtracted) thus reads:
\begin{equation}
  {1\over 2s} \sum_{i\not = j,
    k\not=i,j}F^{n}_{J_1,\ldots,J_n}(\tilde J_1,\ldots,\tilde J_n) 
  {\cal D}_{ij,k} 
  dR_{n+1}(p_1,\ldots,p_{n+1})
\end{equation}
where ${\cal D}_{ij,k} $ denote the dipoles defined in the 
Catani-Seymour subtraction method. Note that the mapping to obtain
the jet momenta $\tilde J_i$ from the parton momenta $p_i$ is encoded in the
dipole. We are not free to chose the mapping in this case as this
would result in a mismatch with the contribution integrated analytically
and combined with the virtual corrections. The contribution from the
subtracted dipoles can thus not be combined point wise with the real
corrections calculated using \Eq{eq:MasterEq}.  In \Ref{Campbell:2013uha} a similar conclusion regarding the
application of Catani-Seymour dipole subtraction within that method is drawn.

So far we have assumed only final state singularities. The above
approach can be easily extended to initial state singularities. The
jet function needs to be extended to cover also initial state
singularities. Using the different mappings as introduced in 
\Refs{Catani:1996vz,Catani:2002hc} is sufficient to handle all
different cases.

\section{Phase space parameterisation}
\label{sec:PhaseSpaceparameterisation}
The parameterisation of the $n+1$ particle phase space in terms of an
$n$ particle phase space times an `unresolved' phase space follows the
phase space factorisation as given in the context of the
Catani-Seymour subtraction method \cite{Catani:1996vz,Catani:2002hc}.
As mentioned before, the mapping of the real phase space to the
reduced kinematics defines the clustering prescription for the
$3\rightarrow 2$ jet algorithm, generalising the $2\to1$ clustering
normally used. Since in the modified jet algorithm the resolution
 is not affected and the recombined jet momenta reproduce the
naive soft and collinear limits\footnote{In the soft limit $(p_i\to
  0)$ the kinematics used in the Catani-Seymour formalism
  reduces to $\{ p_1,\ldots\xcancel{p_i},\ldots,p_{n+1}\}$, while in
  the collinear limit ($p_i\to z p,p_j\to (1-z) p$) the 
  set of momenta reduces to $\{
  p_1,\ldots\xcancel{p_i},\ldots,\xcancel{p_j},\ldots,p_{n+1},p\}$.
  }, the modified jet algorithm automatically fulfills infrared safety,
factorisation of initial state collinear singularities, and momentum
conservation while keeping the resulting jets on-shell. Furthermore,
the phase space factorisation allows to span the respective real phase space associated with each point in
the $n$-jet phase space  in a straightforward
manner.

Each combination of unresolved partons $i,j$ picked by the resolution
 of the jet algorithm
($\Theta(\ycut-\yij\left(p_1,\ldots,p_{n+1}\right))=1$)  
and the spectator $k$ selected through $\Theta_k$ 
defines a specific mapping to cluster $n+1$ partons to $n$ jets
$\left(p_1,\ldots,p_{n+1}\right)\xrightarrow{i,j,k}\left(J_1,\ldots,J_n\right)$.
Depending on whether $j$ and $k$ are final or initial state particles
($i$ is always a final state parton) 
there are four qualitatively different types of mappings which can be
formulated for massless or massive particles \cite{Catani:1996vz,Catani:2002hc}.
To apply the method outlined in section~\ref{sec:formalism} we need to
invert these mappings. For a given set of on-shell jet momenta 
$(J_1,\ldots,J_n)$ and a set of variables describing the unresolved
phase space ($\Phi$) we need the mapping
\begin{equation}
  (J_1,\ldots,J_n, \Phi) \to (p_1,\ldots,p_{n+1})
\end{equation}
to generate the $n+1$ parton phase space. In the following subsections
we collect the required formulae. As mentioned in the introduction
related formulae can be found in 
\Refs{Weinzierl:2001ny,Dinsdale:2007mf,Schumann:2007mg}.

\subsection{Final state clustering with final state spectator}
\subsubsection*{Massless particles}
As described in \Ref{Catani:1996vz} the phase space of $n+1$
massless partons can be factorised in terms of a phase space of $n$
massless momenta---which we identify with the jet momenta $J_i$---and the dipole
phase space measure $dR_{ij,k}$ related to the emission of an
additional parton:
\begin{equation}\label{eq:psff}
  dR_{n+1}\left(p_a+p_b,p_1,\ldots,p_{n-2},p_j,p_k,p_{i}\right)
  =dR_{n}\left(p_a+p_b, J_1,\ldots,J_{n-2},J_{j},J_{k}\right)dR_{ij,k}
\end{equation}
with the $n$-particle phase space as defined in \Eq{eq:phasespace}.
The incoming momenta are given by $p_a$ and $p_b$.
The dipole
phase space measure as given in Ref.~\cite[(5.20)]{Catani:1996vz} reads
in four space time dimensions
\begin{equation}
dR_{ij,k}=\frac{J_j\cdot J_k}{2\left(2\pi\right)^{3}}d\phi\;dz\;dy\left(1-y\right)\Theta\left(\phi\left(2\pi-\phi\right)\right)\Theta\left(z\left(1-z\right)\right)\Theta\left(y\left(1-y\right)\right).
\end{equation}
The set of variables $\Phi = \{\phi,z,y\}$ used to parameterise the phase space
$dR_{ij,k}$ is discussed below.
The phase space parameterisation corresponds to the following
clustering of $n+1$ partons to $n$ jets
\begin{eqnarray}
  J_j &=& p_i+p_j-\frac{y}{1-y}p_k,\\
  J_k &=&\;\frac{1}{1-y}p_k,\\
  J_m &=&\;p_m, \quad\textrm{ for } m\neq j,k
\end{eqnarray}
which fulfill momentum conservation $(\sum\limits_{i=1}^n J_i=P)$ 
and the respective on-shell conditions $(J^2_i=0,\quad i=1,\ldots,n)$.
To invert the mapping $(p_1,\ldots,p_{n+1}) \to (J_1,\ldots,J_n,\Phi)$
we first observe that the momentum $p_k$ and the momenta $p_m$ ($m\neq i,j,k$)
can be obtained in terms of the momenta $J_i$ and the variable $y$ through
\begin{eqnarray}
  p_m &=& J_m,\\
  p_k &=& \left(1-y\right)J_k.
\end{eqnarray}
To determine the missing momenta $p_i$ and $p_j$ we use
\begin{eqnarray}
  p_{ij}&\equiv& p_i+p_j=\;J_j+yJ_k,\\
  s_{ij}&=& \left(J_j+yJ_k\right)^2 = 2y(J_j\cdot J_k).
\end{eqnarray}
The momenta $p_i$ and $p_j$ can be easily expressed in the rest frame of $p_{ij}$ rotated
such that $J_j$ points along the positive $z$-axis. 
Using $2(J_j\cdot p_{ij}) = s_{ij}$ the momentum $J_j'$ in this
particular frame is given by
\begin{equation}
  \label{eq:JinRestFrame}
  J'_j = \frac{\sqrt{s_{ij}}}{2} (1,0,0,1).
\end{equation}
The momentum $J'_j$ is obtained from the
given $J_j$ by a boost into the rest frame of $p_{ij}$ and two
subsequent rotations to annihilate the $x$ and $y$ component of $J_j$:
\begin{equation}
  J'_j = \Lrotx{\phi_x} \Lroty{\theta_y}
  \Lboost{\hat p_{ij}}
  J_j,
\end{equation} 
with the Lorentz transformations $\Lrotx{\phi_x}$,
$\Lroty{\theta_y}$, and 
$\Lboost{\hat p_{ij}}$ given in the appendix. 
We used the hat to denote the parity transform of a four vector $x$:
$\hat x = (x^0,-\vec{x})$.
The angles $\theta_y$ and $\phi_x$ are determined through
\begin{eqnarray}
  \label{eq:thetaangle}
  &&\cos(\theta_y)=\frac{J^z_j}{\sqrt{(J^x_j)^{2}+(J^z_j)^{2}}},\quad 
  \sin(\theta_y)=\frac{J^x_j}{\sqrt{(J^x_j)^{2}+(J^z_j)^{2}}},\\
  \label{eq:phiangle}
  &&\cos(\phi_x)=\frac{\sqrt{(J^x_j)^{2}+(J^z_j)^{2}}}{|\vec{J}_j|},\quad 
  \sin(\phi_x)=\frac{-J^y_j}{|\vec{J}_j|}.
\end{eqnarray}
In this frame the momenta $p'_i$ and $p'_j$ read
\begin{eqnarray}
  \label{eq:iimassless1}
  p'_i &=& {\sqrt{s_{ij}}\over 2} 
  (1,2\sqrt{z(1-z)}\cos\phi,\;2\sqrt{z(1-z)}\sin\phi,\;2z-1),\\
  p'_j &=& \hat p'_i.
\end{eqnarray}
where we have used the definition \cite[(5.6)]{Catani:1996vz}
\begin{equation}
  z = {2 (p_i\cdot J_k)\over 2 (J_j\cdot J_k)}. 
\end{equation}
The momenta $p_i,p_j$ follow from $p'_i,p'_j$ by inverting the Lorentz 
transformations:
\begin{eqnarray}
  \label{eq:iimassless3}
  p_i &=& \Lboost{p_{ij}} \Lroty{-\theta_y} \Lrotx{-\phi_x} p'_i,\\
  \label{eq:iimassless4}
  p_j &=& \Lboost{p_{ij}} \Lroty{-\theta_y} \Lrotx{-\phi_x} p'_j.
\end{eqnarray}
(The inverse of $\Lboost{\hat p_{ij}}$ is given by $\Lboost{p_{ij}}$.)

\subsubsection*{Massive particles}\label{sec:ffm}
For massive partons $i$, $j$, $k$ the phase space can again be factorised
in terms of a phase space of $n$ jets and the dipole phase space 
measure $dR_{ij,k}$ related to the clustered parton \cite{Catani:2002hc}:
\begin{equation}
  dR_{n+1}\left(P,p_1,\ldots,p_{n-2},p_i,p_k,p_{j}\right)
  = dR_{n}\left(P,J_1,\ldots,J_{n-2},J_j,J_{k}\right)dR_{ij,k}.
\end{equation}
The $n$-jet phase space is again given by \Eq{eq:phasespace}, where some
of the $m_i$ are non-zero now. 
The dipole phase space measure as taken
from Ref. \cite[(5.11)]{Catani:2002hc} reads in four dimensions
\begin{eqnarray}
  \nonumber dR_{ij,k} &=& \frac{Q^2}{4\left(2\pi\right)^{3}}
  \frac{\left(1-\mu^2_i-\mu^2_j-\mu^2_k\right)^2}
  {\sqrt{\Kallen{1,\mu^2_{ij},\mu^2_k}}}
  \Theta\left(1-\mu_i-\mu_j-\mu_k\right)\\
  & &\times d\phi\;dz\;dy\left(1-y\right)
  \Theta\left(\phi\left(2\pi-\phi\right)\right)
  \Theta\left(\left(z-z_-\right)\left(z_+-z\right)\right)
  \Theta\left(\left(y-y_-\right)\left(y_+-y\right)\right)
\end{eqnarray}
with the K\"all\'en function defined by
\begin{equation}
\lambda(x,y,z)=x^2+y^2+z^2 - 2xy - 2xz - 2yz,
\end{equation}
and
\begin{equation}
  \mu_n=\frac{m_n}{\sqrt{Q^2}},\quad
  m_{ij}=\sqrt{J^2_j},\quad Q=p_i+p_j+p_k=J_j+J_k.
\end{equation}
The integration boundaries are given by \cite[(5.13)]{Catani:2002hc}
\begin{eqnarray}
  y_-&=&\frac{2\mu_i\mu_j}{1-\mu^2_i-\mu^2_j-\mu^2_k},\\
  y_-&=&1-\frac{2\mu_k\left(1-\mu_k\right)}{1-\mu^2_i-\mu^2_j-\mu^2_k},\\
  z_{\pm}&=&\frac{2\mu^2_i+\left(1-\mu^2_i-\mu^2_j-\mu^2_k\right)y}{2\left[\mu^2_i+\mu^2_j+\left(1-\mu^2_i-\mu^2_j-\mu^2_k\right)y\right]}
  \left(1\pm \varv_{ij,i}\varv_{ij,k}\right),
\end{eqnarray}
with the relative velocities between $p_i+p_j$ and $p_i$ or $p_k$ 
\cite[(5.14)]{Catani:2002hc}
\begin{eqnarray}
  \varv_{ij,i}&=&\frac{\sqrt{\left(1-\mu^2_i-\mu^2_j-\mu^2_k\right)^2y^2-4\mu^2_i\mu^2_j}}{\left(1-\mu^2_i-\mu^2_j-\mu^2_k\right)y+2\mu^2_i},\\
  \varv_{ij,k}&=&\frac{\sqrt{\left[2\mu^2_k+\left(1-\mu^2_i-\mu^2_j-\mu^2_k\right)\left(1-y\right)\right]^2-4\mu^2_k}}{\left(1-\mu^2_i-\mu^2_j-\mu^2_k\right)\left(1-y\right)}.
\end{eqnarray}

The phase space parameterisation corresponds to the following
clustering of $n+1$ partons to $n$ jets \cite[(5.9)]{Catani:2002hc}
\begin{eqnarray}
  J_{k} &=& \sqrt{\frac{\Kallen{1,\mu^2_{ij},\mu^2_k}}
    {\Kallen{1,\frac{s_{ij}}{Q^2},\mu^2_k}}}p_k
  +\left(-\sqrt{\frac{\Kallen{1,\mu^2_{ij},\mu^2_k}}
      {\Kallen{1,\frac{s_{ij}}{Q^2},\mu^2_k}}}\frac{2p_k\cdot
  Q}{Q^2}+\mu^2_k-\mu^2_{ij}+1\right)\frac{Q}{2}\\ 
  J_{j}&=&Q-J_{k},   \\ 
  J_m&=&p_m, \quad (m\neq j,k)
\end{eqnarray}
which again fulfill momentum conservation ($\sum\limits_{i=1}^n J_i=P$) and
and the on-shell conditions ($J^2_j=m^2_{ij},\, J^2_l=m^2_l$ for $l\neq j$).
To invert this clustering the momenta $p_k$ and
$p_m$ ($m\neq i,j,k$) are calculated first:
\begin{eqnarray}
  p_k&=&\left[J_k-\left(1+\mu^2_k-\mu^2_{ij}\right)
    \frac{Q}{2}\right]\sqrt{
    \frac{\Kallen{1,\frac{s_{ij}}{Q^2},\mu^2_k}}
    {\Kallen{1,\mu^2_{ij},\mu^2_k}}}
  +\left[\left(1-y\right)\left(1-\mu^2_i-\mu^2_j-\mu^2_k\right)
    +2\mu^2_k\right]\frac{Q}{2},\\
  p_m&=&J_m,
\end{eqnarray}
where we have used 
\begin{equation}
  2 (p_k\cdot Q) = Q^2+p_k^2-(p_i+p_j)^2 = Q^2 + m_k^2 - s_{ij},
\end{equation}
together with the definition
\begin{equation}
  y = {2 (p_i\cdot p_j)\over Q^2} = {s_{ij} - m_i^2 -m_j^2 \over
    Q^2 - m_i^2-m_j^2-m_k^2}.
\end{equation}
Similar to the massless case it is convenient to express $p_i$ and $p_j$ in the rest frame of $p_{ij}=p_i+p_j$ rotated
such that $Q$ in the respective frame points along the positive
$z$-axis. Using $2(Q\cdot p_{ij}) = Q^2 + s_{ij}-m_k^2$  the momentum
$Q'$ in this particular frame is then given by
\begin{equation}
  Q'=\frac{Q^2}{2\sqrt{s_{ij}}}\left(\frac{s_{ij}}{Q^2}+1-\mu^2_k,0,0,
    \sqrt{\Kallen{1,\frac{s_{ij}}{Q^2},\mu^2_k}}\right).
\end{equation}
Again $Q'$ is obtained from $Q$ through a boost to the
rest frame of $p_{ij}$ and subsequent rotations:
\begin{equation}
  Q'= \Lrotx{\phi_x} \Lroty{\theta_y} \Lboost{\hat p_{ij}} Q,
\end{equation} 
where the angles are similar to \Eq{eq:thetaangle} and \Eq{eq:phiangle}.
In this frame the momenta $p'_i$ and $p'_j$ read
\begin{eqnarray}
  p'_i&=&\left(\frac{Q^2}{2\sqrt{s_{ij}}}
  \left(\frac{s_{ij}}{Q^2}+\mu^2_i-\mu^2_j\right),
  \left|\vec{p}'_i\right| (\sin\theta'\cos\phi,\;\sin\theta'\sin\phi,
    \;\cos\theta') \right),\\
  p'_j&=&\left( \frac{Q^2}{2\sqrt{s_{ij}}}
  \left(\frac{s_{ij}}{Q^2}+\mu^2_j-\mu^2_i\right), -\vec{p}'_i \right),
\end{eqnarray}
with $  \left|\vec{p}'_i\right|=\sqrt{(p'^0_i)^2-m^2_i} = {1\over 2
  \sqrt{s_{ij}}} \sqrt{\Kallen{s_{ij},m_i^2,m_j^2}}$.
Using
\begin{eqnarray}
  s_{ik}&=&\left(p_i+p_k\right)^2=\;Q^2\left[z\left(1-y\right)
    \left(1-\mu^2_i-\mu^2_j-\mu^2_k\right)+\mu^2_i+\mu^2_k\right]
\end{eqnarray}
one gets
\begin{equation}
  \cos{\theta'} = {Q^2 (1-y)\*(1-\mu_i^2-\mu_j^2-\mu_k^2)\*
 [ ((1-\mu_i^2-\mu_j^2-\mu_k^2) y 
   + \mu_i^2+\mu_j^2 ) (1-2 z)-\mu_j^2+\mu_i^2]\over
   \sqrt{\Kallen{s_{ij},m_i^2,m_j^2}}
     \sqrt{\Kallen{1,\frac{s_{ij}}{Q^2},\mu^2_k}}}.
\end{equation}
Under the exchange $\mu_i^2 \leftrightarrow \mu_j^2, z \to 1-z$ we have
$\cos(\theta') \to - \cos(\theta')$ as it should be.

The momenta $p_i,p_j$ follow again from $p'_i,p'_j$ by
\begin{eqnarray} 
  p_i &=& \Lboost{p_{ij}} \Lroty{-\theta_y} \Lrotx{-\phi_x} p'_i,\\
  p_j &=& \Lboost{p_{ij}} \Lroty{-\theta_y} \Lrotx{-\phi_x} p'_j.
\end{eqnarray}

\subsection{Final state clustering with initial state spectator}
\subsubsection*{Massless particles}\label{ssec:fi}
Using Ref.~\cite[(5.45)]{Catani:1996vz} the phase space can be factorised
into the phase space of $n$ massless particles and the
dipole phase space $dR_{ij,a}$: 
\begin{equation}
\label{eq:FImassive}
dR_{n+1}\left(p_a+p_b, p_1,\ldots,p_{n-1},p_j,p_i\right)
=dR_{n}\left(xp_a+p_b,J_1,\ldots,J_{n-1},J_j\right)dR_{ij,a}.
\end{equation}
The dipole phase space measure in four dimensions is given by
\cite[(5.48)]{Catani:1996vz}
\begin{equation}
dR_{ij,a}=\frac{J_j\cdot p_a}{2(2\pi)^3}d\phi\;dz\;dx\;\Theta\left(\phi\left(2\pi-\phi\right)\right)\Theta\left(z\left(1-z\right)\right)\Theta\left(x\left(1-x\right)\right).
\end{equation}
Note that $dR_{ij,a}$ includes an integration over $x$ leading to a
convolution of the measures given in \Eq{eq:FImassive}.
The space parameterisation corresponds to the
clustering of $n+1$ partons to $n$ jets
\begin{eqnarray}
  \label{eq:masslessMappingija1}
  J_j&=&\;p_i+p_j-\left(1-x\right)p_a,\\
  \label{eq:masslessMappingija2}
  J_m&=&\;p_m\quad (m\neq i,j)
\end{eqnarray}
which fulfills momentum conservation ($\sum\limits_{i=1}^n J_i= x p_a+ p_b$)
and the on-shell conditions ($J^2_l=0$, $l=1,\ldots,n$).
Inverting this clustering allows to parameterise $n+1$ partons by
means of the $n$ jet momenta and three integration variables
$x,z,\phi$ as follows
\begin{equation}
p_m=J_m, \quad (m\neq i,j)
\end{equation}
and because of
\begin{eqnarray}
p_{ij}&=&p_i+p_j=\;J_j+\left(1-x\right)p_a,\\
s_{ij}&=&\left(J_j+\left(1-x\right)p_a\right)^2,
\end{eqnarray}
$p_i$ and $p_j$ can be calculated using the steps outlined in 
\Eq{eq:JinRestFrame} -- \Eq{eq:iimassless4}.

\subsubsection*{Massive particles}\label{ssec:fim}
Using \Ref{Catani:2002hc} the phase space of $n+1$ massive partons can be
expressed as a phase space of $n$ particles convoluted with the dipole
phase space $dR_{ij,a}$:
\begin{equation}
  dR_{n+1}\left(p_a+p_b, p_1,\ldots,p_{n-1},p_j,p_i\right)
  =dR_{n}\left(xp_a+p_b, J_1,\ldots,J_{n-1},J_j\right)dR_{ij,a}.
\end{equation}
The dipole phase space measure $dR_{ij,a}$ reads \cite[(5.48)]{Catani:2002hc}:
\begin{equation}
dR_{ij,a}=\frac{J_j\cdot p_a}{2\left(2\pi^3\right)}d\phi\;dz\;dx\;\Theta\left(\phi\left(2\pi-\phi\right)\right)\Theta\left(\left(z-z_-\right)\left(z_+-z\right)\right)\Theta\left(x\left(x_+-x\right)\right).
\end{equation}

The integration boundaries are given by
\begin{eqnarray}
x_+&=&1+\mu^2_{ij}-\left(\mu_i+\mu_j\right)^2,\\
z_{\pm}&=&\frac{1-x+\mu^2_{ij}+\mu^2_i-\mu^2_j\pm\sqrt{\left(1-x+\mu^2_{ij}-\mu^2_i-\mu^2_j\right)^2-4\mu^2_i\mu^2_j}}{2\left(1-x+\mu^2_{ij}\right)}.
\end{eqnarray}
with
\begin{equation}
\mu_n=\frac{m_n}{\sqrt{2J_j\cdot p_a}},\quad
m_{ij}=\sqrt{J^2_j}.
\end{equation}
The phase space parameterisation corresponds to the clustering of
$n+1$ partons to $n$ jets as in \Eq{eq:masslessMappingija1} and 
\Eq{eq:masslessMappingija2}
but satisfying now the on-shell conditions
$J^2_j=m^2_{ij}$ and $J^2_l=m^2_l$ for $l\neq j$.
To invert the mapping $(p_1,\ldots,p_{n+1}) \to
(J_1,\ldots,J_n,\Phi)$ ($\Phi=\{x,z,\phi\}$), we start again in the
rest frame of $p_{ij}=p_i+p_j$ rotated such, that the momentum $J_j$
points along the positive $z$-axis.
Using $(p_{ij}\cdot J_j) = {1\over 2}(s_{ij}-m_{ij}^2)$ the corresponding
momenta $J'_j$ in the rest frame of $p_{ij}$ is given by
\begin{equation}
  J'_j = \frac{1}{2\sqrt{s_{ij}}}\left(s_{ij}+m_{ij}^2,
    0,0,s_{ij}-m_{ij}^2\right).
\end{equation}
The relation to $J_j$ is again given by a sequence of one Lorentz
boost and two rotations: 
\begin{equation}
\label{eq:Jtransformation}
J'_j = \Lrotx{\phi_x} \Lroty{\theta_y} \Lboost{\hat p_{ij}}J_j .
\end{equation} 
the momenta $p'_i$ and $p'_j$ are given by
\begin{eqnarray}
  p'_i &=& \left(\frac{s_{ij}-m^2_j+m^2_i}{2\sqrt{s_{ij}}},
    |\vec{p}'_i|\left(\sin\theta'\cos\phi,\;\sin\theta'\sin\phi,\;\cos\theta'\right)\right),\\
  p'_j & = & \left(\frac{s_{ij}-m^2_i+m^2_j}{2\sqrt{s_{ij}}},-\vec{p}'_i\right),
\end{eqnarray}
with $\left|\vec{p}'_i\right|=\sqrt{{E'_i}^2-m^2_i}={1\over
  2\sqrt{s_ij}} \sqrt{\Kallen{s_{ij},m_i^2,m_j^2}}$ and
\begin{align}
 \cos{\theta'}
=&\frac{m^2_j-m^2_i - (1-2z)s_{ij}}{
  \sqrt{\Kallen{s_{ij},m_i^2,m_j^2}}
}.
\end{align}

$p_i$ and $p_j$ follow from $p'_i$ and $p'_j$ by inverting the Lorentz
transformations given in \Eq{eq:Jtransformation}.

\subsection{Initial state clustering with final state spectator }
\subsubsection*{Massless particles}
The phase space of $n+1$ massless partons can be expressed as a
phase space convolution of a phase space of $n$ massless jets and the
dipole phase space measure $dR_{ia,k}$ for the emission of an
additional massless parton from the initial state with a massless
final state spectator. Most statements from section \ref{ssec:fi} can
be carried over by the replacements $a\rightarrow k$ and $j\rightarrow
a$ (see Ref. \cite{Catani:1996vz}). However, since we now are dealing
with clustering in the initial state, collinear singularities must be
factorisable into the parton distribution functions.  Because
\begin{subequations}\label{eq:colfacif}
\begin{align}
x=\frac{p_{a}\cdot\left(p_{i}+p_{k}\right)-p_{i}\cdot p_{k}}{p_{a}\cdot\left(p_{i}+p_{k}\right)}\xrightarrow[p_i\rightarrow (1-z)p_a]{}z,
\end{align}
a jet function applying this clustering fulfills the condition for factorisability of initial state collinear singularities 
\begin{align}
F^{n+1}_{J_1,\ldots,J_n}\left(p_1,\ldots,p_{n-1},p_k,p_i;p_a,p_b\right)\xrightarrow[p_i\rightarrow (1-z)p_a]{}F^{n}_{J_1,\ldots,J_n}\left(p_1,\ldots,p_{n-1},p_k;zp_a,p_b\right).
\end{align}
\end{subequations}

\subsubsection*{Massive particles}
The phase space of $n$ massive partons and one massless parton
can be expressed as a phase space
convolution of a phase space of $n$ massive jets and the dipole phase
space measure $dR_{ia,k}$ for the emission of an additional massless
parton from the initial state with a massive final state spectator.
All statements from section \ref{ssec:fim} can be carried over by the
replacements $a\rightarrow k$ and $j\rightarrow a$, $m_i\rightarrow 0$
and $m_{ij}\rightarrow m_k$ (see \Ref{Catani:2002hc}). The
argument from \Eq{eq:colfacif} also holds.

\subsection{Initial state clustering with initial state spectator}\label{sec:ii}

In case of initial state clustering with an initial state spectator
the phase space can again be written as a convolution 
\cite[(5.149)]{Catani:1996vz}:
\begin{equation}\label{eq:psii}
  dR_{n+1}\left(p_a+p_b,p_1,\ldots,p_n,p_i\right)
  =dR_{n}\left(xp_a+p_b, J_1,\ldots,J_{n}\right)dR_{ia,b},
\end{equation}
with the $n$-particle phase space given in \Eq{eq:phasespace}.
The dipole phase space measure $dR_{ia,b}$ reads 
\cite[(5.151)]{Catani:1996vz}:
\begin{align}
  dR_{ia,b}=\frac{p_a\cdot p_b}{2\left(2\pi\right)^3}
  d\phi\;d\varv\;dx\;\Theta\left(\phi\left(2\pi-\phi\right)\right)
  \Theta\left(\varv\right)\Theta\left(1-\frac{\varv}{1-x}\right)
  \Theta\left(x\left(1-x\right)\right).
\end{align}

The phase space parameterisation corresponds to the following
clustering of $n+1$ (massless/massive) partons to $n$ 
(massless/massive) jets:
\begin{equation}
J_m=\;{\Lambda_{ia,b}}\;p_m, \quad m=1,\ldots,n
\end{equation}
with
\begin{equation}
K=\;p_a+p_b-p_i,\quad \widetilde{K}=\;xp_{a}+p_b,
\end{equation}
and the Lorentz boost transforming $K$ into $\widetilde K$ given by
\cite[(5.144)]{Catani:1996vz}:
\begin{eqnarray}\label{eq:ltrsfii}
  {\left[\Lambda_{ia,b}\right]^{\mu}}_{\nu}&=& {g^{\mu}}_{\nu}
  -\frac{2\left(K+\widetilde{K}\right)^{\mu}\left(K+\widetilde{K}\right)_{\nu}}
  {\left(K+\widetilde{K}\right)^2}
  +\frac{2\widetilde{K}^{\mu}K_{\nu}}{K^2}.
\end{eqnarray}
The inverse boost is obtained by exchanging $K$ and $\widetilde K$.
All outgoing momenta $p_i$ are transformed to balance the transverse
momentum. Momentum conservation ($\sum\limits_{i=1}^n J_i=x p_a+p_b$)
and on-shell conditions ($J^2_l=m^2_l$, $l=1,\ldots,n$) are not
affected by the boost.
Inverting this clustering allows to parameterise $n+1$ partons by
means of the $n$ jet momenta and three integration variables
$x,\varv,\phi$ as follows. The momenta $p_m$ $(m=1,\ldots,n$) are
obtained by inverting the boost:
\begin{equation}
  p_m=\Lambda_{ia,b}^{-1}\;J_m.
\end{equation}
Using the definition for $\varv$ 
\begin{equation}
   s_{ia}=\;\varv s_{ab},  
\end{equation}
together with
\begin{equation}
  s_{ib}=\left(1-x-\varv\right)s_{ab}
\end{equation}
which leads to
\begin{equation}
  s_{ia}+s_{ib} = (1-x)s_{ab}
\end{equation}
it is straightforward to express the momentum $p_i$ in the rest frame
of $p_a+p_b$ rotated such that $p_a$ points along the $z$-axis. In
this particular frame the momentum $p_i$ is given by
\begin{equation}
  p'_i=\left(1-x\right)\frac{\sqrt{s_{ab}}}{2}(1,\sin\theta'_i\cos\phi,
  \;\sin\theta'_i\sin\phi,\;\cos\theta'_i).
\end{equation}
Using $p_a$ in this particular frame
\begin{equation}
  p'_a = \frac{\sqrt{s_{ab}}}{2}\left(1,0,0,1\right),\\
\end{equation} 
the angle $\theta'$ can be read off
\begin{equation}
\cos\theta'_i=1-\frac{2\varv}{1-x}.
\end{equation}

The momenta $p_i$ is obtained according to
\begin{equation}
  p_i=\Lroty{-\theta_y}\Lrotx{-\phi_x}p'_i
\end{equation}
as in the previous cases with $J_j \to p_a$. Note that no massive case as in the previous
sections needs to be
studied since the two incoming partons and the
collinear parton are always assumed to be massless.

\section{Consistency checks}\label{sec:checks}
To validate the approach, we apply the procedure outlined in the
previous sections to two processes. As an example, where one has to
deal with initial state singularities, we study Drell-Yan production
in hadronic collisions.  More precisely, we calculate Drell-Yan
production at the LHC running at a center-of-mass energy of
$\sqrt{s}=13$~TeV. We apply phase space cuts similar to what is used
in the LHC experiments. For simplicity, we veto any additional jet in
the final state since we are only interested in the case where
recombination occurs in the real corrections. The inclusion of the
contribution due to an additional jet is straightforward since no
recombination occurs.  For the final state electrons, we require the
invariant mass $m_{ee}$ of the electron pair in the region defined by
$116$ GeV$<m_{ee}<3$ TeV.  Furthermore, we demand a minimum transverse
momentum $p^{\perp}_{e}>25$ GeV for the electron and restrict the
rapidity of the electron to $|\eta_e|<2.5$.  Unresolved initial state
radiation is clustered with
the beam according to section \ref{sec:ii}.\\
As a second example, where one has to deal with final state radiation,
we analyse top-quark pair production in $e^+e^-$ annihilation. Similar
to the Drell-Yan case, we veto again the emission of an additional
jet.  For the top-quark mass we use $m_t=174$~GeV. For the
center-of-mass energy we choose $\sqrt{s}=500$~GeV relevant for a
future linear collider. We do not include the decay of the top quarks,
instead, we treat them as tagged top-jets. These jets are obtained
with a $kt$-jet algorithm with the resolution criteria defined by
\begin{equation}
  \yij=2\;\frac{\min\left(E^2_i,E^2_j\right)
    \left(1 - \cos(\theta_{ij})\right)}{s},
\end{equation}
and the resolution \ycut\ set to $\ycut=0.1$. For the recombination
of unresolved particles the modified $3\rightarrow 2$ clustering
prescription according to section \ref{sec:ffm} is used.

Although very simple, these two examples cover essentially all relevant
cases. Furthermore, compact analytic results are available for the
higher order corrections and it is straightforward to apply the ideas
outlined in this article. For details on the NLO calculations using
phase space slicing we refer to \Refs{Brandenburg:1998xw,Harris:2001sx}.

\begin{figure}[htbp]
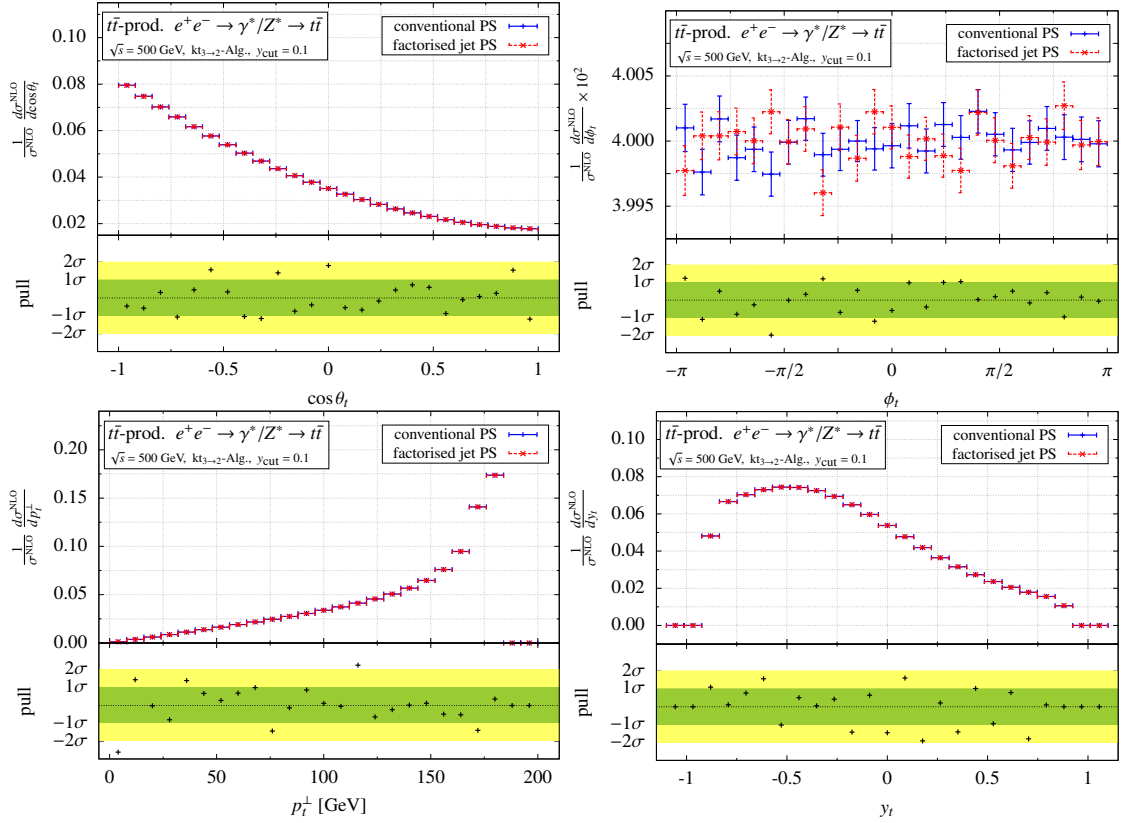

  \begin{center}
    \leavevmode
    \includegraphics[width=0.48\textwidth]{{{%
          eeQQKT3-2ycut0.1compCTH1-50x1e7-crop}}}
    \includegraphics[width=0.48\textwidth]{{{%
          eeQQKT3-2ycut0.1compPH1-50x1e7-crop}}}
    
    \includegraphics[width=0.48\textwidth]{{{%
          eeQQKT3-2ycut0.1compPTR1-50x1e7-crop}}}
    \includegraphics[width=0.48\textwidth]{{{%
          eeQQKT3-2ycut0.1compY1-50x1e7-crop}}}
    \caption{Differential distributions for top-quark pair production 
      in $e^+e^-$ annihilation calculated using a conventional parton
      level MC compared with a calculation using the factorised jet
      phase space as described in section \ref{sec:ffm}.}
    \label{fig:partonMCvsjetMC_ttprod}
  \end{center}
\end{figure}

Exclusively demanding $n$ jets in the final state allows to define a
differential $n$-jet event weight at NLO accuracy:
\begin{equation}\label{eq:diffwgt}
  \frac{d\sigma^{\NLO}}{d^4J_1\dots d^4J_{n}}
  = \frac{d\sigma^{\textrm{B}}}{d^4J_1\dots d^4J_{n}}
  + \frac{d\sigma^{\textrm{V}}}{d^4J_1\dots d^4J_{n}}
  + \frac{d\sigma^{\textrm{R}}}{d^4J_1\dots d^4J_{n}}.
\end{equation}
We use the superscripts B, V and R to indicate the contributions from
the Born matrix elements, the virtual corrections and the real
corrections. In case of the real corrections a regularisation of the
soft- and collinear singularities using the phase space slicing method
is understood. The `unresolved' contribution is included in the
virtual corrections and cancels the respective soft and collinear
singularities. We note, that the real corrections are calculated using
\Eq{eq:MasterEq} which means that for each phase space point
$(J_1,\ldots,J_n)$ an additional three dimensional integration is
required to obtain $d\sigma^{\textrm{R}}$. To check the approach and
the numerical implementation we  use \Eq{eq:diffwgt} integrated over
the phase space to calculate
the total cross section. The results can be compared with the ones
obtained by a standard parton level Monte Carlo.  We have checked that
the results using \Eq{eq:MasterEq} are in perfect agreement
with the results of a conventional parton MC. 
One may argue that the comparison of the total cross
section is not very sensitive to the details of the calculation and
inconsistencies in specific phase space regions could escape
detection.  In fact \Eq{eq:diffwgt} can also be used to calculate
arbitrary distributions:
\begin{equation}\label{eq:jetobs}
  \frac{d\sigma^\NLO}{dO(\widetilde{J}_1,...,\widetilde{J}_{n})}\\
  =\int\prod\limits_{m=1}^{n}d^4J_m
  \frac{d\sigma^\NLO}{d^4J_1\dots d^4J_{n}}
  \delta\left(O(J_1,...,J_n)-O(\widetilde{J}_1,...,\widetilde{J}_n)\right).
\end{equation}
Again these contributions can be compared with the outcome of
a parton level Monte Carlo. This comparison allows a detailed
check of the entire phase space. We stress that in the parton level
Monte Carlo the same modified jet algorithm ($3\rightarrow 2$
clustering!) has to be used. 
\begin{figure}[htbp]
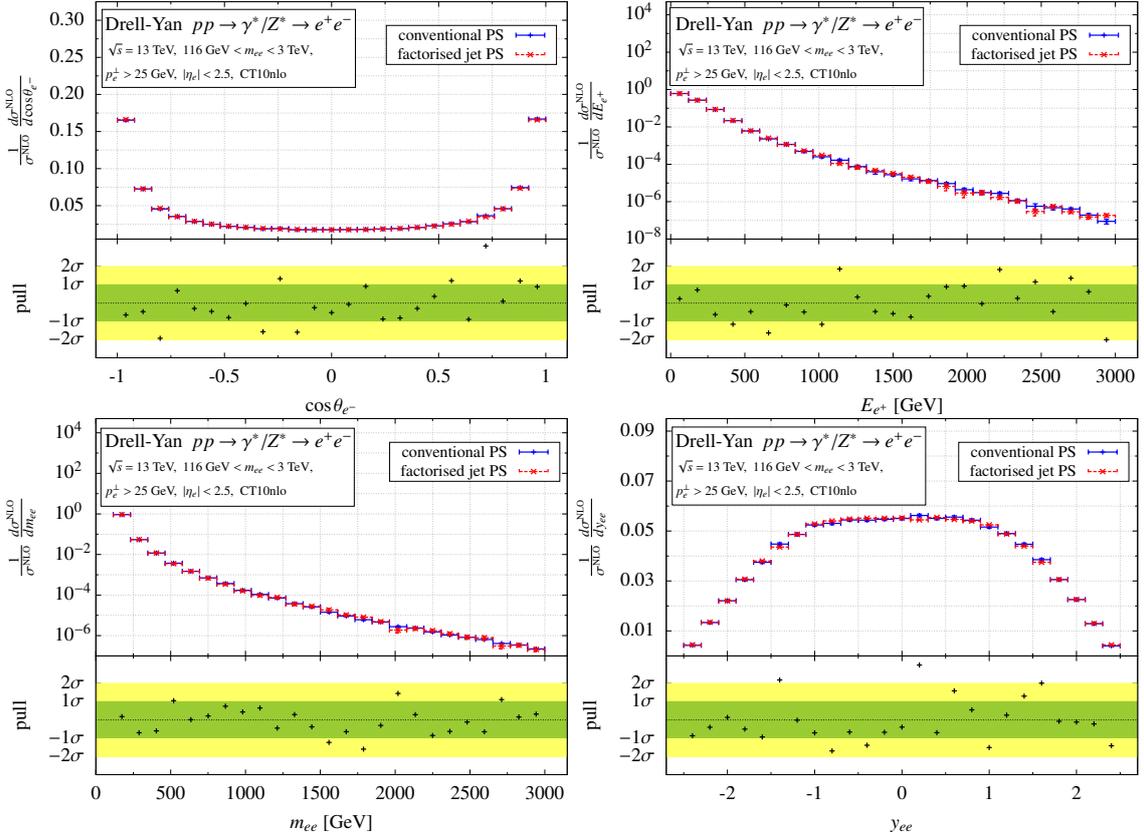

  \begin{center}
    \leavevmode
    \includegraphics[width=0.49\textwidth]{{{%
          qqee3-2ptmn25etamx2.5compCTH2-50x1e7-crop}}}
    \includegraphics[width=0.49\textwidth]{{{%
          qqee3-2ptmn25etamx2.5compE1-50x1e7-crop}}}
    
    \includegraphics[width=0.49\textwidth]{{{%
          qqee3-2ptmn25etamx2.5compMEE-50x1e7-crop}}}
    \includegraphics[width=0.49\textwidth]{{{%
          qqee3-2ptmn25etamx2.5compYEE-50x1e7-crop}}}
    \caption{Differential distributions for Drell-Yan production
      calculated using a conventional parton level MC compared with a
      calculation using the factorised jet phase space as described in
      section \ref{sec:ii}.}
    \label{fig:partonMCvsjetMC_DrellYan}
  \end{center}
\end{figure}
In \Fig{fig:partonMCvsjetMC_DrellYan} we collect various distributions
calculated for Drell-Yan production. In particular, we show the
angular and the energy distribution of the scattered electron/positron. In
addition, the invariant mass distribution and the rapidity distribution
of the $e^+e^-$ system are given. The blue solid lines show the results
obtained with a conventional parton level Monte Carlo. The red dashed
lines show the results using the factorised jet phase space as
illustrated in the previous sections. In the lower part of the plots
we show the discrepancy between the two approaches in terms of
standard deviations, where the uncertainty of each approach is due to
the limited statistics of the Monte Carlo integrations. For the parton
distribution functions we use the
CT10nlo pdf set \cite{Lai:2010vv}. The center-of-mass energy is set
to 13~TeV and we applied the aforementioned cuts. 
For all four distributions, we find
perfect agreement between the two approaches. In most cases the
discrepancy is less than one standard deviation. In
\Fig{fig:partonMCvsjetMC_ttprod} a similar analysis is shown for
top-quark pair production in $e^+e^-$ annihilation. In particular, we
show distributions with respect to the cosine of the azimuthal angle
of the outgoing top quark, the polar angle distribution, the
transverse momentum distribution and the rapidity distribution. Again
the blue solid curves show the results of a conventional parton level
Monte Carlo while the red dashed curves give the results using the
factorised jet phase space. Note that in both cases the modified $3\to
2$ clustering is employed. Again we find perfect agreement between the
two approaches.
\subsection{Impact of NLO corrections --- $k$-factors}
It has been pointed out in \Ref{Campbell:2012cz} that restricting the NLO
analysis to the Born level kinematics may lead to rather
moderate $k$-factors in general .
 \begin{figure}[htbp]
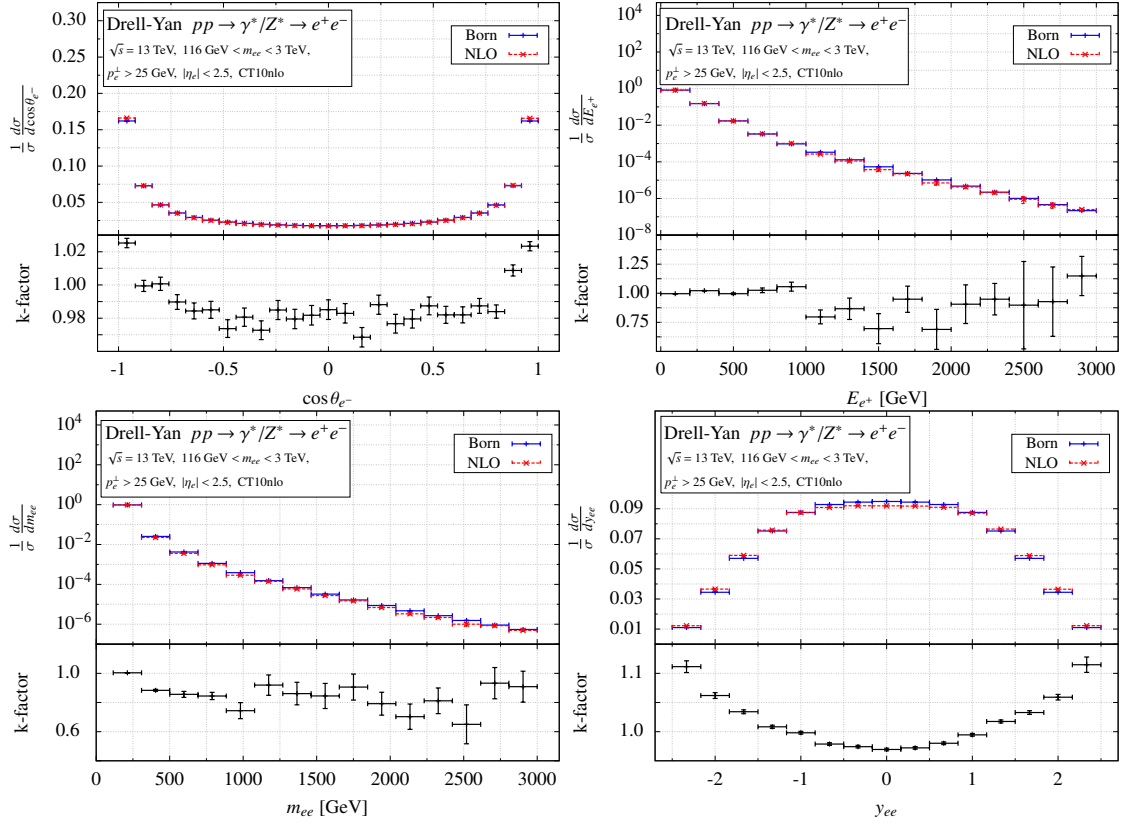

  \begin{center}
    \leavevmode
    \includegraphics[width=0.48\textwidth]{{{%
          qqee3-2ptmn25etamx2.5compbornCTH2-50x1e7-crop}}}
    \includegraphics[width=0.48\textwidth]{{{%
          qqee3-2ptmn25etamx2.5compbornE1-100x1e7-crop}}}

    \includegraphics[width=0.48\textwidth]{{{%
          qqee3-2ptmn25etamx2.5compbornMEE-100x1e7-crop}}}
    \includegraphics[width=0.48\textwidth]{{{%
          qqee3-2ptmn25etamx2.5compbornYEE-100x1e7-crop}}}
    \caption{Impact of NLO corrections on differential distributions 
      for Drell-Yan production at a hadron collider.}
    \label{fig:KFactor_dy}
  \end{center}
\end{figure}
In \Fig{fig:KFactor_dy} we show the respective $k$-factors  for the previously studied
differential distributions. In case of the
angular distribution of the outgoing electron, NLO corrections at the
level of only a few percent are observed. For the rapidity
distributions they are slightly larger but still small in absolute
size. For the energy distribution of the electron and the invariant
mass distribution of the lepton pair the corrections seem to be
larger. However, the $k$-factor suffers from statistical uncertainties
and shows large fluctuations. In regions where the statistical
fluctuations are small we find again a moderate $k$-factor. 
\begin{figure}[htbp]
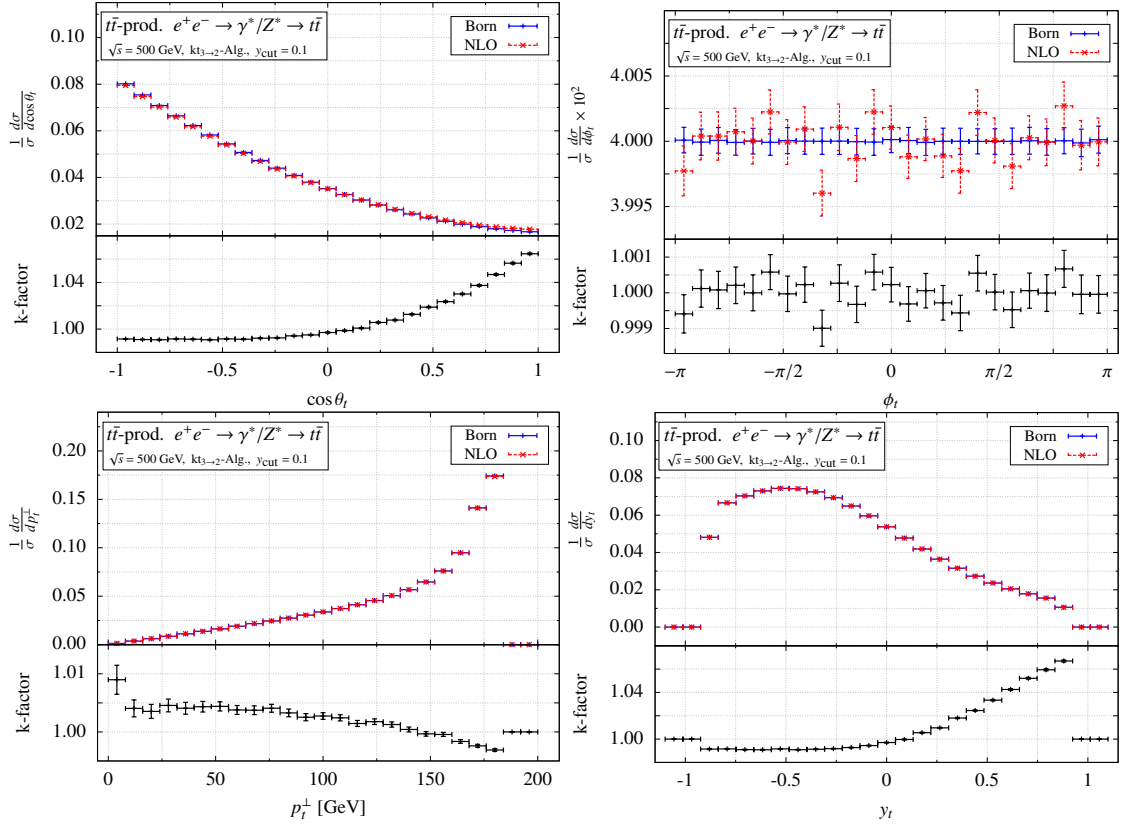

  \begin{center}
    \leavevmode
    \includegraphics[width=0.48\textwidth]{{{%
          eeQQKT3-2ycut0.1compbornCTH1-50x1e7-crop}}}
    \includegraphics[width=0.48\textwidth]{{{%
          eeQQKT3-2ycut0.1compbornPH1-50x1e7-crop}}}

    \includegraphics[width=0.48\textwidth]{{{%
          eeQQKT3-2ycut0.1compbornPTR1-50x1e7-crop}}}
    \includegraphics[width=0.48\textwidth]{{{%
          eeQQKT3-2ycut0.1compbornY1-50x1e7-crop}}}
    \caption{Impact of NLO corrections on differential distributions
      for top-quark pair production in $e^+e^-$ annihilation.}
    \label{fig:KFactor_tt}
  \end{center}
\end{figure}
In \Fig{fig:KFactor_tt} the $k$-factor for top-quark pair production
in $e^+e^-$ annihilation is shown. In all cases we find NLO
corrections of a few per cent only and thus $k$-factors very close to one. 
We thus extend the observations of \Ref{Campbell:2012cz} also to final state 
radiation.

\subsection{Impact of modified clustering / jet algorithms}
All the previously shown differential distributions have been obtained
using the modified jet algorithm: in the conventional parton level Monte
Carlo as well as in the alternative approach using a factorised jet
phase space. As pointed out in the introduction we consider the
modification of the clustering as part of the intrinsic ambiguities of
jet algorithms. As a consequence we do not expect a large effect of
the clustering. If in contrast a large effect is observed one should
question the definition of the observable since it shows a large
sensitivity on an aspect of the jet algorithm which is not well
defined. Since it is difficult to make general statements about the
size of possible effects, one should investigate the impact of the 
new clustering on a case by case study to make sure that no large
deviations are observed.
\begin{figure}[htbp]
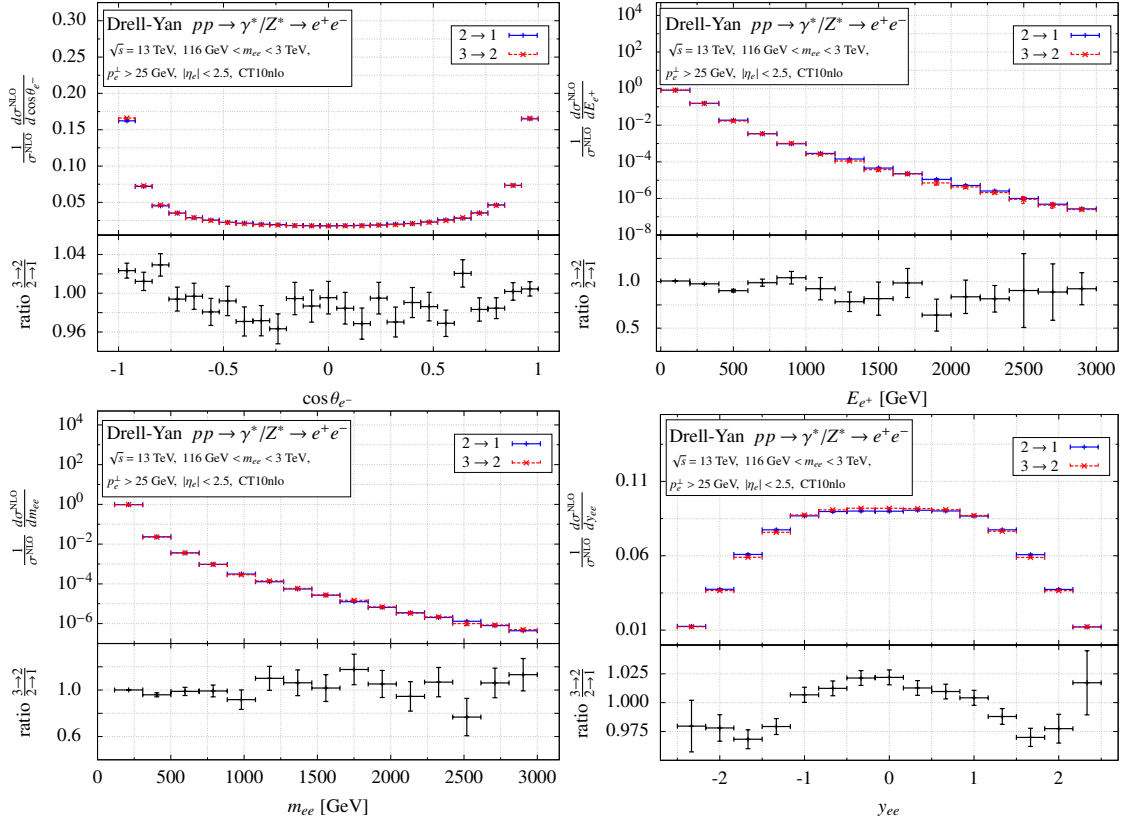

  \begin{center}
    \leavevmode
    \includegraphics[width=0.48\textwidth]{{{%
          qqee3-2ptmn25etamx2.5compnloCTH2-50x1e7-crop}}}
    \includegraphics[width=0.48\textwidth]{{{%
          qqee3-2ptmn25etamx2.5compnloE1-100x1e7-crop}}}

    \includegraphics[width=0.48\textwidth]{{{%
          qqee3-2ptmn25etamx2.5compnloMEE-100x1e7-crop}}}
    \includegraphics[width=0.48\textwidth]{{{%
          qqee3-2ptmn25etamx2.5compnloYEE-100x1e7-crop}}}
    
    \caption{Impact of $3\rightarrow 2$ clustering with respect to 
      $2\rightarrow 1$ clustering on differential distributions for
      Drell-Yan production at a hadron collider.}
    \label{fig:imp3221_dy}
  \end{center}
\end{figure}
In \Fig{fig:imp3221_dy} the conventional $2\to 1$ clustering which
does not respect the on-shell condition of the clustered objects is compared
for Drell-Yan production
with the $3\to2$ clustering advocated here. Note that both results
have been obtained using a conventional parton-level MC. The blue
solid curves show the result using the $2\to1$ clustering, while the
red dashed lines give the results for the $3\to2$ clustering. 
Since for Drell-Yan production the clustering never includes the
outgoing electron we do not expect a major effect. Indeed
\Fig{fig:imp3221_dy} shows essentially no difference within the
statistical uncertainty. A minor effect is visible in the angular
distribution and in the rapidity distribution. This can be related to
the initial state clustering which may introduce an additional boost
orthogonal to the beam axis which can influence the angular distributions.
\begin{figure}[htbp]
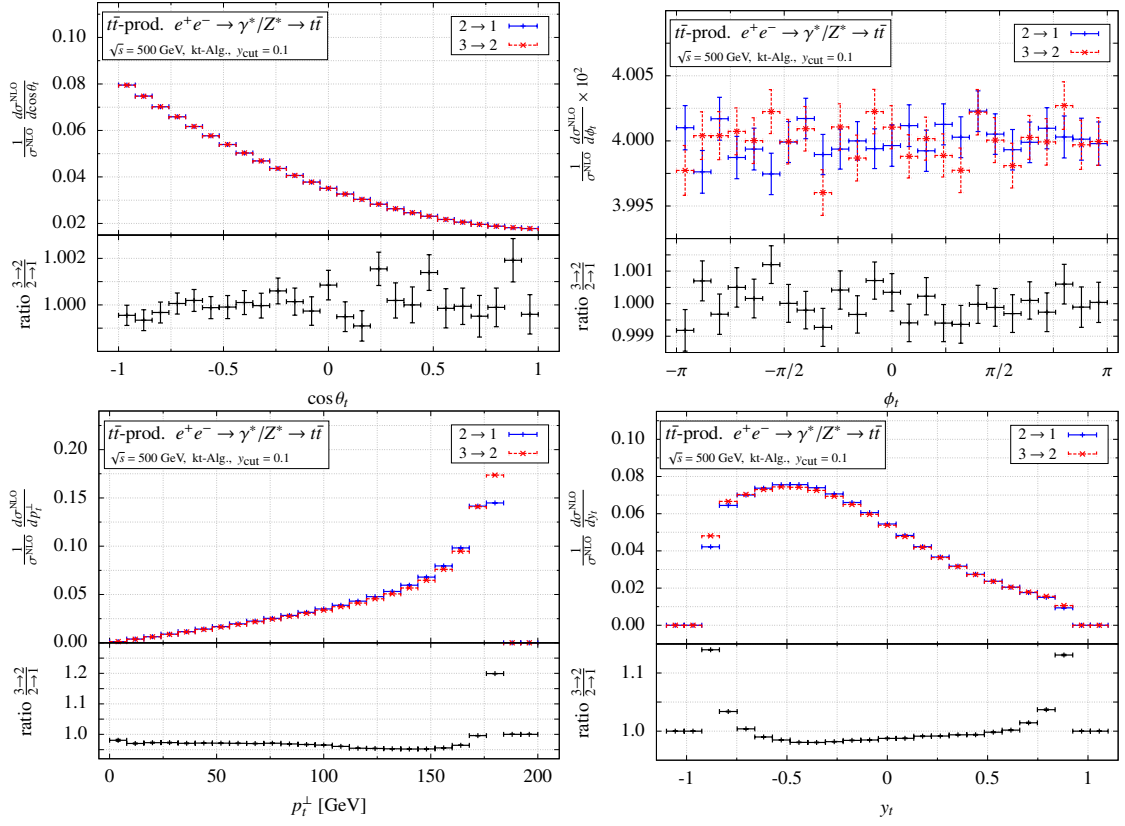

  \begin{center}
    \leavevmode
    \includegraphics[width=0.48\textwidth]{{{%
          eeQQKT3-2ycut0.1compnloCTH1-50x1e7-crop}}}
    \includegraphics[width=0.48\textwidth]{{{%
          eeQQKT3-2ycut0.1compnloPH1-50x1e7-crop}}}

    \includegraphics[width=0.48\textwidth]{{{%
          eeQQKT3-2ycut0.1compnloPTR1-50x1e7-crop}}}
    \includegraphics[width=0.48\textwidth]{{{%
          eeQQKT3-2ycut0.1compnloY1-50x1e7-crop}}}
    
    \caption{Impact of $3\rightarrow 2$ clustering with respect to 
      $2\rightarrow 1$ clustering on differential distributions for
      top-quark pair production in $e^+e^-$ annihilation.}
    \label{fig:imp3221_tt}
  \end{center}
\end{figure}
In \Fig{fig:imp3221_tt} the corresponding result is shown for
top-quark pair production. For the angular distribution the two
different algorithms give the same result within the statistical 
uncertainties. For the transverse momentum distribution a large effect
is visible at large transverse momentum. This is not surprising since
at phase space boundaries we expect to become sensitive to the details
of the clustering. Below 160 GeV we observe that the $3 \rightarrow 2$ clustering
leads to distributions which are between two and five per cent below
the traditional $2 \rightarrow 1$ combination. In general the $2\to1$ clustering
leads to an increase of the mass of the clustered object which could
be responsible for the observed pattern. 
\begin{figure}[htbp]
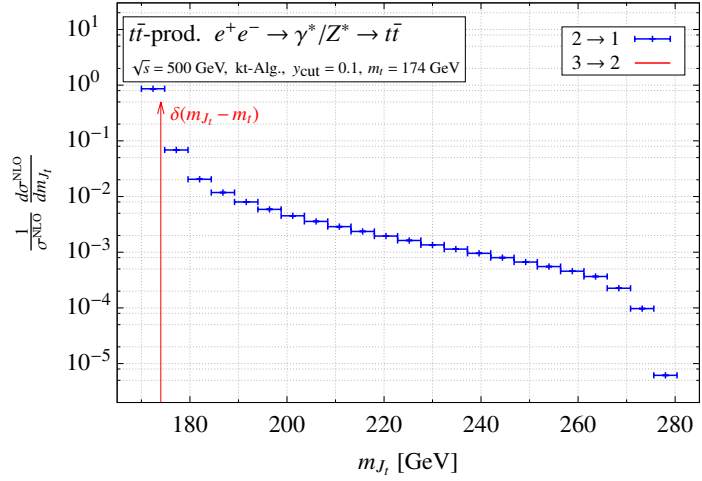

  \begin{center}
    \leavevmode
    \includegraphics[width=0.6\textwidth]{{{%
          eeQQKT3-2ycut0.1konvM1-50x1e7_21-crop}}}
    \caption{Distribution of the mass of the top-quark jet using the
      conventional $2\to1$ recombination.}
    \label{fig:massdistr_2to1_clustering}
  \end{center}
\end{figure}
To analyse this effect we show in \Fig{fig:massdistr_2to1_clustering}
the distribution of the mass of the top-quark jet using the
conventional $2\to 1$ clustering. As one can see most of the events
have a jet-mass close to the nominal top-quark mass. However, there
are also events with jet masses up to 280 GeV. Note that using
the modified clustering the jet mass is fixed to the top-quark
mass. In particular at phase space boundaries, the difference in the
jet mass may result in distortions of distributions which are
sensitive to mass effects. This is precisely what we observe in 
the lower plots of \Fig{fig:imp3221_tt} where one can see that indeed
the largest effects arise at the phase space boundary. This is not
surprising since minor changes in the mass of the clustered objects
become important. Since the distributions are normalised this effect
introduces also a modification in the distribution away from the phase
space boundary. Using cuts to
avoid the phase space boundaries should thus result in smaller
differences between the two different clustering prescriptions. This
is illustrated in \Fig{fig:imp3221_tt_cuts} where we show the same
distributions but now using additional cuts.
\begin{figure}[htbp]
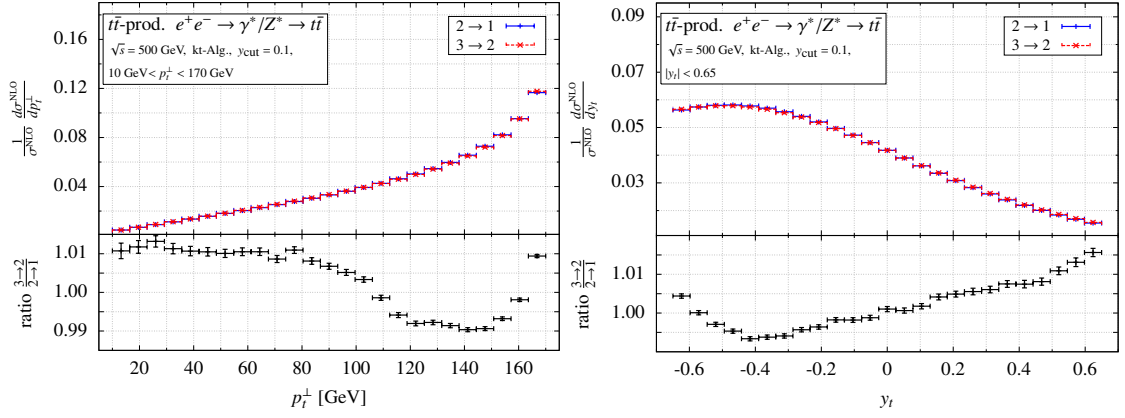

  \begin{center}
    \leavevmode
   \includegraphics[width=0.48\textwidth]{{{%
          eeQQKT3-2ycut0.1compnloPTR1-50x1e7_cuts2-crop}}}
    \includegraphics[width=0.48\textwidth]{{{%
          eeQQKT3-2ycut0.1compnloY1-50x1e7_cuts2-crop}}}
    
    \caption{Same as \Fig{fig:imp3221_tt} but with additional cuts to
      avoid the phase space boundaries.}
    \label{fig:imp3221_tt_cuts}
  \end{center}
\end{figure}
Again the blue solid line shows the conventional $2\to1$ clustering
while the red dashed line shows the alternative $3\to2$ clustering.
Indeed we find that the difference becomes smaller and is of the order
of $1 \%$ only, which might be seen as an intrinsic uncertainty.
To close this section we
stress that minor differences between the two clusterings are not per
se problematic as long as everything is done consistently and the same
clustering is used in the experimental analysis.

\section{Application}\label{sec:application}
In this section we apply the MEM to top-quark pair production in
$e^+e^-$ annihilation. First
we use the aforementioned procedure to generate unweighted events with
NLO accuracy. These events are then analysed using the MEM in LO and
NLO accuracy. In particular, we illustrate the extraction of the
top-quark mass from the event sample.

\subsection{Generating unweighted jet events}
For a chosen jet algorithm with a preset value of the resolution \ycut\ it
is now straightforward to generate unweighted jet events using an 
`acceptance-rejection' algorithm. The respective NLO jet weight is given by
\begin{equation}\label{eq:evwgtn}
  \rho\left(J_1,...,J_n\right)=
  \frac{d\sigma^{\NLO}}{d^4J_1\dots d^4J_{n}}.
\end{equation}
where the right hand side is evaluated according to \Eq{eq:diffwgt}.
The acceptance-rejection method requires an upper boundary
$\rho_{\textrm{max}}$ of the weight
$\rho\left(J_1,...,J_n\right)$. This can be obtained for example within
a phase space integration. An $n$-jet candidate
event is then constructed using $(3n-4)$ random numbers. As a measure
for the probability the weight introduced in \Eq{eq:evwgtn} is
calculated for the candidate event. Note that a
three dimensional integration must be performed to do so. Generating an
additional uniformly distributed random number $r_{\textrm{u}}$ between 0 and 
$\rho_{\textrm{max}}$ the candidate event is accepted if
$r_{\textrm{u}}$ is below the aforementioned weight.\\
 In principle it is also possible to generate unweighted NLO $n$-jet events
$(J_1,...,J_n)$ together with $n+1$-jet events $(J'_1,...,J'_{n+1})$
from $n+1$ partons $(p_1,...,p_{n+1})$ by augmenting the definition
of $\rho$:
\begin{eqnarray}\label{eq:evwgtn+1}
  \nonumber\tilde\rho\left(p_1,...,p_{n+1}\right)
  &=& \frac{d\sigma^{\NLO}}{
    d^4J_1\dots d^4J_{n}}F^{n+1}_{J_1,\ldots,J_n}(p_1,...,p_{n+1})\\
  & &+\frac{d\sigma^{\NLO}}{
    d^4J'_1\dots d^4J'_{n+1}}F^{n+1}_{J'_1,\ldots,J'_{n+1}}(p_1,...,p_{n+1}).
\end{eqnarray}
The jet functions $F^{n+1}_{J_1,\ldots,J_n}$ and
$F^{n+1}_{J'_1,\ldots,J'_{n+1}}$ decide whether $n$ or $n+1$ jets
are resolved and how the momenta $p_i$ are clustered into the jets.
The $n+1$-jet events $(J'_1,...,J'_{n+1})$ are obtained by the
identification
\begin{equation}\label{eq:n+1jet}
J'_i\equiv p_i.
\end{equation}
The $n$-jet events $(J_1,...,J_n)$ follow from clustering by the
$3\rightarrow 2$ jet algorithm
\begin{equation}\label{eq:njetclus}
  (p_1,...,p_{n+1})\rightarrow  
  \left(J_1(p_1,...,p_{n+1}),...,J_n(p_1,...,p_{n+1})\right).
\end{equation}
The main difference with respect to the previously described event
generation is, that now $n+1$ parton momenta are generated using
$(3(n+1)-4)$ random numbers. While this method will generated $n$-jet
events with NLO accuracy we stress that the generated $n+1$-jet events have only
LO accuracy. 

\begin{figure}[htbp]
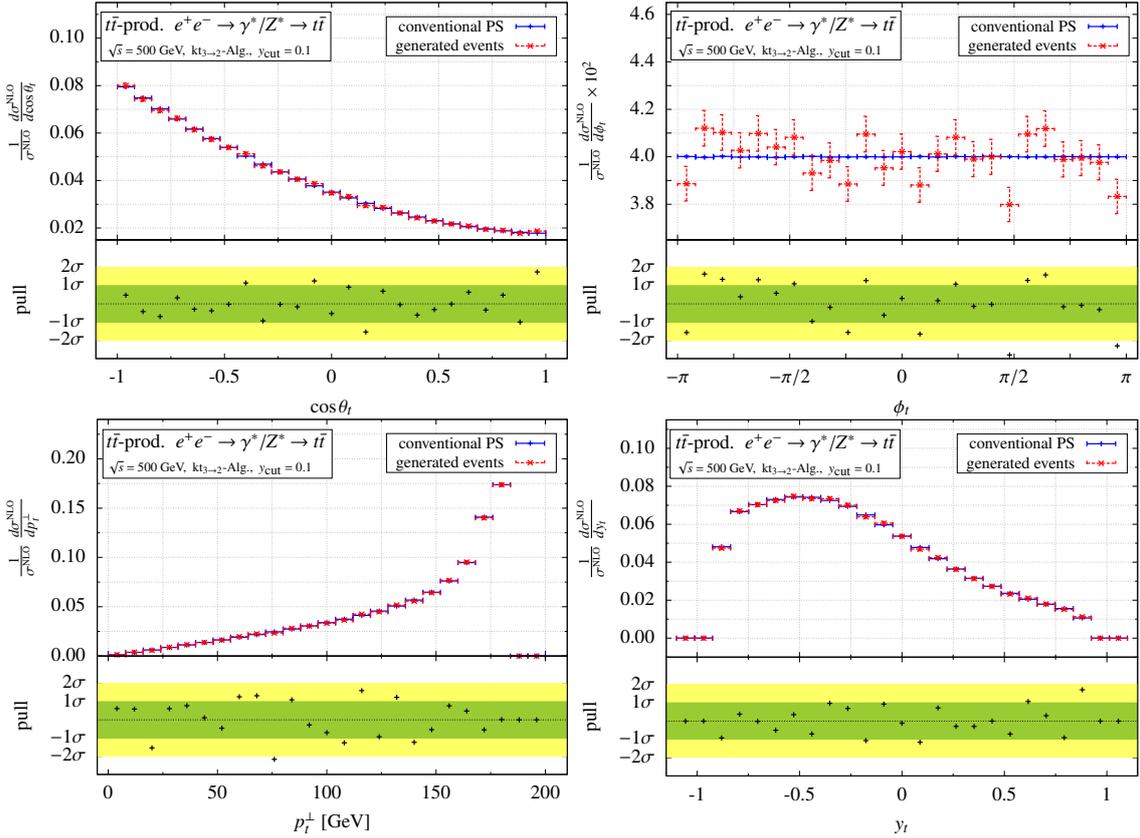

  \begin{center}
    \leavevmode
    \includegraphics[width=0.49\textwidth]{{{%
          eeQQKT3-2ycut0.1compCTH1-150evts-crop}}}
    \includegraphics[width=0.49\textwidth]{{{%
          eeQQKT3-2ycut0.1compPH1-150evts-crop}}}

    \includegraphics[width=0.49\textwidth]{{{%
          eeQQKT3-2ycut0.1compPTR1-150evts-crop}}}
    \includegraphics[width=0.49\textwidth]{{{%
          eeQQKT3-2ycut0.1compY1-150evts-crop}}}
    
    \caption{Validation of the generation of unweighted NLO 
      top-quark pair events (no additional jet)}
    \label{fig:partonMCvsunwgtEv_ttprod}
  \end{center}
\end{figure}

To validate the generation of unweighted events, we reproduce the 
differential distributions calculated in section \ref{sec:checks}.
In total we generated $73128$ events with NLO accuracy. As in section
\ref{sec:checks} we veto the emission of an additional jet. 
In \Fig{fig:partonMCvsunwgtEv_ttprod} we show the comparison with
distributions calculated using the conventional parton level Monte Carlo.
The blue solid lines represent the results from the parton-level Monte
Carlo while the red dashed lines show the distributions calculated
from the unweighted jet events generated as described above. Below we
show the difference between the two distribution in units of one
standard deviation. As one can see we find perfect agreement within
the statistical uncertainties.

\subsection{\MEM}
The possibility to generate unweighted jet events with NLO accuracy 
together with the possibility to assign NLO event weights to them,
allows to perform a validation of the \MEM at NLO
using the generated events as input to a toy experiment. 
As a concrete example we illustrate the extraction of the top-quark mass 
in $e^+e^-\rightarrow t\bar{t}$ employing the
MEM at NLO. We note that this study  may be relevant for the
top-quark mass measurements at a future linear collider.

\begin{figure}[htbp]
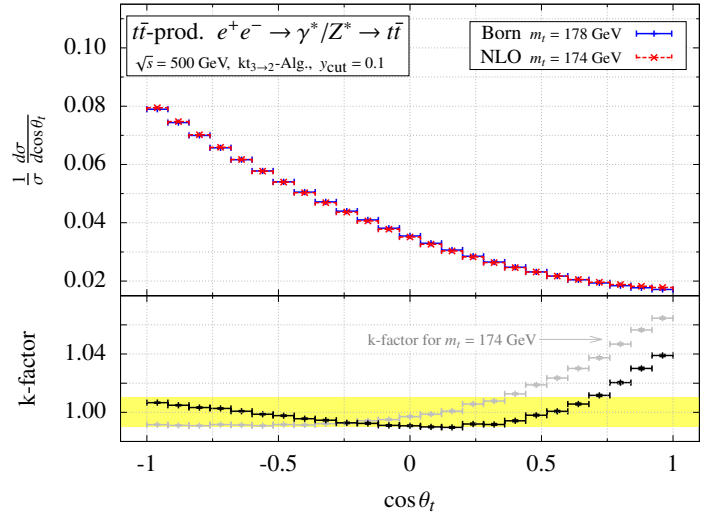

  \begin{center}
    \leavevmode
    \includegraphics[width=0.6\textwidth]{{{%
          eeQQKT3-2ycut0.1compbornCTH1-50x1e7mt178-crop}}}
    \caption{Comparison of the NLO predictions (red dashed line) evaluated for $\mt=174$~GeV with the Born approximation (blue solid line) evaluated for
      $\mt=178$~GeV. In the lower plot the ratio of the two is shown (black line). For comparison the k-factor for $m_t=174$~GeV is shown in gray (cf. \Fig{fig:KFactor_tt}).}
    \label{fig:born178vsNLO174}
  \end{center}
\end{figure}

As mentioned before we ignore for simplicity the top-quark decay and
assume that top-(anti)quark jets are observed. 
An event is than defined by the energies and angles
of the respective jets $(E_t,\;\cos{\theta_{t}},\;\phi_{t},\;E_{\bar{t}},
\;\cos{\theta_{\bar{t}}},\;\phi_{\bar{t}})$. This fixes the jet momenta
depending on the top quark mass $m_t$ as
\begin{eqnarray}\label{eq:jmommt}
\nonumber 
J_{t}&=&\left(E_{t},\;|p_t|\cos{\phi_{t}}\sin{\theta_{t}},
  \;|p_t|\sin{\phi_{t}}\sin{\theta_{t}},
  \;|p_t|\cos{\theta_{t}}\right),\\
J_{\bar{t}} &=& \left(E_{\bar{t}},
  \;|p_{\bar t}|\cos{\phi_{\bar{t}}}\sin{\theta_{\bar{t}}},
  \;|p_{\bar t}|\sin{\phi_{\bar{t}}}\sin{\theta_{\bar{t}}},
  \;|p_{\bar t}|\cos{\theta_{\bar{t}}}\right),
\end{eqnarray}
with $|p_{t,\bar t}| = \sqrt{E_{t,\bar t}^2-m_t^2}$. 
Exclusively demanding a top-quark pair without an additional jet fixes 
the energies $E_t=E_{\bar{t}}=\sqrt{s}/2$. 
A 2-jet NLO event weight for $\vec{x}=(\cos{\theta_{t}},\;\phi_{t},
\;\cos{\theta_{\bar{t}}},\;\phi_{\bar{t}})$ can be obtained according to
\begin{eqnarray}
  \nonumber\frac{d\sigma^\NLO}{d\vec{x}}
  &=&\frac{d\sigma^\NLO}{d\cos{\theta_{t}}
    \;d\phi_{t}\;d\cos{\theta_{\bar{t}}}\;d\phi_{\bar{t}}}\\ \nonumber
  &=&
  \frac{\beta_t}{32\pi^2}\;\frac{d\sigma^\NLO}
  {d^4J_{1}\;d^4J_{2}}\bigg|_{\cos{\theta_{1}}=\cos{\theta_{t}},
    \;\cos{\theta_{2}}=\cos{\theta_{\bar{t}}},
    \;\phi_{1}=\phi_{t},\;\phi_{2}=\phi_{\bar{t}}}.
\end{eqnarray}
with $\beta_t = \sqrt{1-\frac{4m^2_t}{s}}$.
A sample of $N$ unweighted 2-jet NLO events 
\begin{equation}
\left\{\vec{x}_i=(\cos{\theta^i_{t}},\;\phi^i_{t},\;\cos{\theta^i_{\bar{t}}},\;\phi^i_{\bar{t}}),\;i=1,...,N\right\}
\end{equation}
is generated for some `true" top-quark mass
$m^\true_t=174\;\textrm{GeV}$.  The NLO likelihood
$\mathcal{L}^{NLO}$ for the sample can be constructed from the
differential 2-jet cross section as follows
\begin{eqnarray}
  \nonumber\mathcal{L}^\NLO\left(m_t\right)
  &=& 
  \prod\limits_{i=1}^{N} \mathcal{L}^\NLO
  \left(\vec{x}_i|m_t\right)=\left(\frac{1}{\sigma^\NLO(m_t)}\right)^N
  \prod\limits_{i=1}^{N}\frac{d\sigma^\NLO(m_t)}{d\vec{x}_i}\\
  &=& \left(\frac{\beta_t}{32\pi^2\;\sigma^\NLO(m_t)}\right)^N
  \prod\limits_{i=1}^{N}
  \left.\left(\frac{d\sigma^\NLO}{d^4J_{t}\;d^4J_{\bar{t}}}
    \right)\right|_{\mbox{\scriptsize event }i}(\mt)
\end{eqnarray}
where the dependence on $m_t$ is shown explicitly and $J_{t}$ and 
$J_{\bar{t}}$ follow from $\vec{x}_i$ according to
\Eq{eq:jmommt}. Note that the jet momenta when evaluated for the event 
$\vec{x}_i$ depend on the mass $\mt$.
The negative logarithm of the likelihood (or "Log-likelihood") therefore reads
\begin{eqnarray}
  \nonumber
  -\log{\mathcal{L}^\NLO\left(m_t\right)}
  &=&-\sum\limits_{i=1}^{N}\log\mathcal{L}^\NLO
  \left(\vec{x}_i|m_t\right)
  = N\log\left(\sigma^\NLO(m_t)\right)
  -\sum\limits_{i=1}^{N}\log\left(\frac{d\sigma^\NLO(m_t)}{d\vec{x}_i}
  \right)\\
  &=& 
  N\log\left(\frac{32\pi^2\;\sigma^\NLO(m_t)}{\beta_t}\right)
  - \sum\limits_{i=1}^{N}
  \left.\log\left(\frac{d\sigma^\NLO(m_t)}{d^4J_{t}\;d^4J_{\bar{t}}}
    \right)\right|_{\mbox{\scriptsize event }i}.
\end{eqnarray}
Maximising (minimising) this likelihood (Log-likelihood) with respect
to $m_t$ yields an estimator $\widehat{m}_t$
\begin{eqnarray}
  \nonumber\mathcal{L}^{\NLO}\left(\widehat{m}_t\right)
  &=&\sup\limits_{m_t}\left(\mathcal{L}^{\NLO}\left(m_t\right)\right),\\
  -\log\mathcal{L}^{\NLO}\left(\widehat{m}_t\right)
  &=&\inf\limits_{m_t}\left(-\log\mathcal{L}^{\NLO}\left(m_t\right)\right).
\end{eqnarray}
\begin{figure}[htbp]
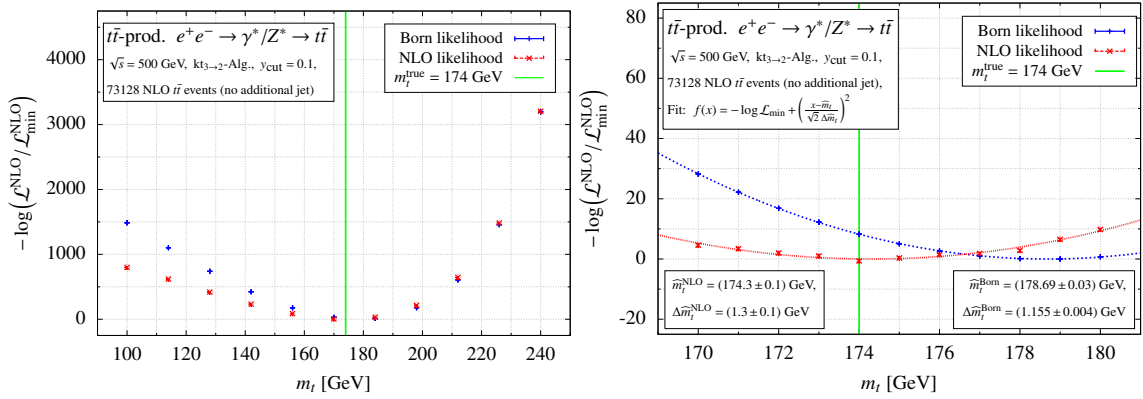

  \begin{center}
    \leavevmode
    \includegraphics[width=0.49\textwidth]{{{%
          eeQQKT3-2ycut01mem-150evts-crop}}}    
    \includegraphics[width=0.49\textwidth]{{{%
          eeQQKT3-2ycut01memdet-150evts-crop}}}
    \caption{NLO and Born Log-likelihood for 73128 NLO events between 
      $m_t=100$ GeV and $m_t=240$ GeV ("true" value
      $m^\textrm{true}_t=174\;\textrm{GeV}$) and zoomed in between
      $m_t=170$ GeV and $m_t=180$ GeV to extract $\widehat{m}_t$ and
      $\Delta\widehat{m}_t$ by a fit}
 \label{fig:LogLmemLONLOhistat}
  \end{center}
\end{figure}
The lefthand plot of \Fig{fig:LogLmemLONLOhistat} illustrates the
Log-likelihood for 73128 events generated with NLO accuracy as
function of the top-quark mass $m_t$ in the range between $100$ GeV and 
$240$ GeV. The solid blue line shows the
result evaluating the likelihood using LO predictions, the dashed red
line shows the result obtained using NLO predictions. As mentioned
before a top-quark mass $m_t^{\true} = 174$~GeV has been used to generate
the events. The righthand plot of \Fig{fig:LogLmemLONLOhistat} shows the
area around the minimum between $m_t=170$ GeV and $m_t=180$ GeV in
order to extract $\widehat{m}_t$ and $\Delta\widehat{m}_t$ by a
parabola fit (see \Ref{cowan1998statistical}).
As one can see from \Fig{fig:LogLmemLONLOhistat}
extracting the top-quark mass with a likelihood based on the Born
approximation yields an estimator $\widehat{m}^\textrm{Born}_t$ 
which shows a significant deviation from the input 
value $m^\true_t$ hence can not be regarded as an unbiased estimator. 
More precisely we find  
\begin{equation}
  \widehat{m}^\textrm{LO}_t = (178.7\pm1.2)\mbox{~GeV}
\end{equation}
which is 4 $\sigma$ way from the true top-quark mass used in the event 
generation. Note that in LO the results are independent of
$\alpha_s$. It is thus not possible to attribute a theoretical
uncertainty by simply varying the renormalisation scale.
On the other hand extracting the
top-quark mass with the likelihood based on NLO predictions results in an
estimator $\widehat{m}^\NLO_t$ which is consistent with the
input value $m^\true_t$ within the uncertainty. Using NLO accuracy we
find
\begin{equation}
    \widehat{m}^\NLO_t = (174.3\pm1.3)\mbox{~GeV}
\end{equation}
in perfect agreement with $m^\true_t$.
We stress that in both cases we use the same unweighted events 
generated with NLO accuracy. 
In principle the discrepancy between $\widehat{m}^\textrm{LO}_t$ and 
$\widehat{m}^\NLO_t$ is not surprising since we used in 
$\widehat{m}^\textrm{LO}_t$ LO predictions to analyse events generated
with NLO accuracy. What is however remarkable is the large size of the
effect. As has been illustrated in section \ref{sec:checks} the NLO
corrections are usually small for most of the
distributions. Nevertheless we observe a large effect using LO or NLO
predictions within the \MEM. From the above results we
may conclude that the Born matrix element evaluated for $\mt=178$ GeV
gives a better approximation of the NLO corrections evaluated for
$\mt=174$~GeV than the Born approximation evaluated for 174~GeV. To
investigate this point further we show in \Fig{fig:born178vsNLO174}
the comparison of the two predictions.  
Obviously the NLO corrections cannot be completely absorbed by changing
the mass in the LO predictions. Comparing however the black with the gray line in
\Fig{fig:born178vsNLO174} we find that indeed
$\mt=178$ GeV gives a slightly better description of the NLO
result. In the range $-1<\cos\theta_t\lesssim0.75$ the difference is below
$1\%$ and the maximal deviation at $\cos\theta_t=1$ is $4\%$. The difference is below $1\%$ in the range $-1<\cos\theta_t\lesssim0.38$ and the maximal deviation at $\cos\theta_t=1$ is $6\%$ when $\mt=174$ GeV is used in the Born approximation.

In view of a future linear collider we
stress that the renormalisation scheme is well defined in the above
procedure. Applying the above procedure to realistic data, the
top-quark mass within the pole mass scheme would be determined.

\begin{figure}[htbp]
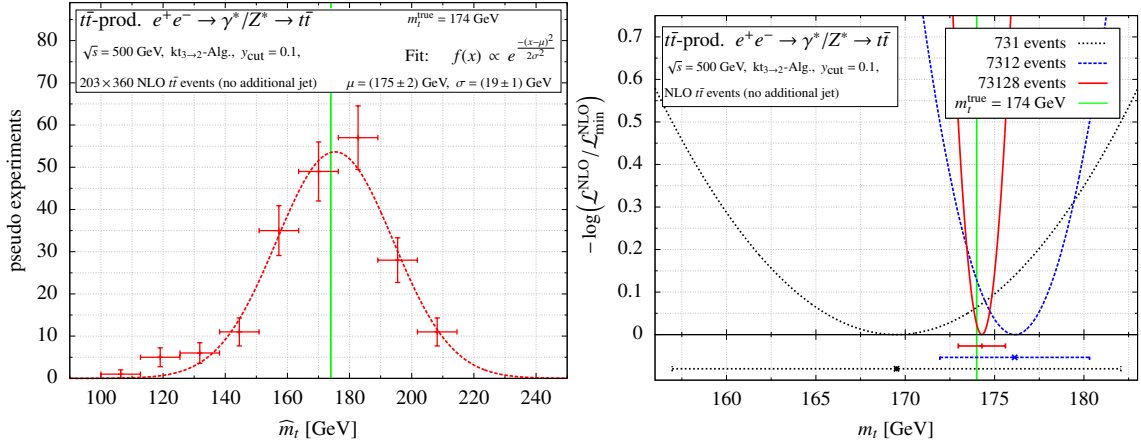

  \begin{center}
    \leavevmode
    \includegraphics[width=0.49\textwidth]{{{%
          eeQQKT3-2ycut01mem-203x360-crop}}}
    \includegraphics[width=0.49\textwidth]{{{%
          eeQQKT3-2ycut01mem-73128-7312-731evts-crop}}}    
    \caption{Distribution of the estimator $\widehat{m}_t$ around the 
      true value $m^\textrm{true}_t=174$ GeV and dependence of the
      Log-likelihood on the number of events. At the bottom of the right plot the respective estimator $\widehat{m}_t\pm\Delta\widehat{m}_t$ is shown.}
    \label{fig:LogLmemNLOhistat}
  \end{center}
\end{figure}
In \Fig{fig:LogLmemNLOhistat} we show the consistency of the
approach. In particular, we illustrate that $\widehat{m}^\NLO_t$
provides indeed an unbiased maximum likelihood estimator
(\Ref{Hoadley:1971}).  The lefthand plot of \Fig{fig:LogLmemNLOhistat}
shows the distribution of the estimator $\widehat{m}^\NLO_t$
if we interpret our event sample with 73128 events as 203 independent
toy experiments with 360 events each. 
The dashed line shows a gaussian fitted to the data. 
The righthand plot illustrates how increasing the number of events results
in a tightening of the dip in the Log-likelihood around the true value
of $m_t$ and the approximate scaling of the error of the estimator $\Delta\widehat{m}_t\propto N^{-\frac{1}{2}}$ (see bottom of righthand plot). 

As a final remark we comment on the impact of the modified
jet algorithm. Top-quark pair production in $e^+e^-$ annihilation
is highly constrained through
momentum conservation and the underlying symmetries of the
interaction. Most of the sensitivity to the top-quark mass stems
essentially from information contained already in in the
$\cos(\theta_t)$ distribution. On the other hand for this distribution
the two clustering algorithms give the same result at the permille
level as we have shown in \Fig{fig:imp3221_tt}. As a consequence we do
not expect major differences in case the conventional clustering would
be used in the experimental analysis.

\section{Conclusion}
\label{sec:conclusion}
In this article we have shown how to calculate event weights for jet
events at NLO
accuracy. The ability to define event weights at NLO is a necessary
prerequisite to extend the \MEM beyond the Born
approximation. The basic ingredient of the method presented here is a
modification of the clustering prescription used in jet algorithms.
Instead of using the conventional $2\to 1$ clustering, where the
momentum of the clustered object is just the sum of the two initial
jet candidates, we use a recombination inspired by the phase space
mapping used in the Catani-Seymour subtraction method. This leads
naturally to a factorisation of the phase space for the real
corrections into resolved and unresolved contributions. Furthermore,
the factorisation allows to integrate numerically the contribution of
the unresolved configurations after an appropriate regulator to handle
the mass and soft singularities has been chosen. 
Similar ideas have been investigated in
a different context already in
\Refs{Weinzierl:2001ny,Dinsdale:2007mf,%
  Schumann:2007mg}. The major difference is
that we consider the new clustering not only as a technical
trick but as an essential part of a modified jet algorithm. Using this
modified clustering no artifical jet mass is generated and it is
straightforward to map the events obtained onto the born kinematics.
As validation of the proposed method, we have successfully reproduced
differential distributions at NLO accuracy in Drell-Yan production and
top-quark pair production in $e^+e^-$ annihilation. Although simple,
these two examples cover essentially all relevant cases.  We have also
investigated the impact of the modified jet algorithm. At phase space
boundaries the effects can be large.  Additional cuts can be used to
reduce the impact. The remaining effect may be considered as an intrinsic
uncertainty inherent to jet algorithms. 
For the examples studied here the effect is reduced to the per cent level,
after applying cuts. We stress however that in hadronic collisions the
situation could be different and one needs to investigate the impact of the
new clustering on a case by case study.
As a further application we
have studied as a toy example the \MEM at NLO applied
to top-quark pair production in $e^+e^-$ annihilation. More precisely,
we investigated the determination of the top-quark mass. This study is
relevant for a possible future Linear Collider.  Applying the \MEM
to events generated with NLO accuracy we observe that
the MEM in LO fails to reproduce the input value. While the NLO
analysis correctly reproduces the input with an uncertainty of
about 1 GeV for about 70000 simulated events, the LO analysis leads to
a value off by 4 GeV. These findings should be taken into account, when
top-quark mass measurements at the Tevatron using the MEM are discussed.
Let us end with a final remark concerning parton shower
corrections. As mentioned in section \ref{sec:formalism} the naive
inclusion of corrections due to the parton shower would lead to a
double counting. Further studies are required to extend the method
presented here in this direction.

\subsection*{Note added:}
While we were in the process of writing this article
\Ref{Giele:2015sva} appeared, where an extension of the jet algorithm
in $e^+e^-$ annihilation to massless quarks similar to what is
discussed here has been presented.

\acknowledgments
We would like to thank Simone Aioli, Fabrizio Caola, Markus Schulze, Thomas Lohse, Oliver Kind, Patrick Rieck and Sören Stamm for useful discussions and Walter Giele and Stefan Weinzierl for a
careful reading of the manuscript. This project has been supported by the 
German Ministry of Education and Research (BMBF) through the BMBF
grant 05H12KHE. PU would like to thank the CERN theory unit, where some
of the reported work has been done, for its hospitality.

\appendix
\section{Explicit form of the Lorentz transformations}
For a given four vector $X$ with $X^2\not=0$ the rotational free
boost from the rest frame of $X$ to the system where $X$ takes the
form as given reads when applied to the four momentum $y$:
\begin{equation}
  \label{eq:BoostRestframe}
  \Lboost{X} y = \left( {X^0\over \sqrt{X^2}}y^0
    +{(\vec{X}\cdot \vec{y})\over \sqrt{X^2}}, 
    \vec{y} 
    + \left[{(\vec{X}\cdot \vec{y})\over \sqrt{X^2}(X^0+\sqrt{X^2})}
    + {y^0\over \sqrt{X^2}}
    \right] \vec{X}\right).
\end{equation}
Defining $\hat X = (X^0,-\vec{X})$ the boost from the frame in which
$X$ is given to the rest frame is given by $\Lambda^b(\hat{X})$.
In fact, \Eq{eq:BoostRestframe} is a special case of the more general boost
given in \Eq{eq:ltrsfii}:
\begin{equation}
  \Lboost{X} = 
  \Lambda_{ia,b},\quad\mbox{for } K=(\sqrt{X^2},\vec{0}),\; 
  \widetilde{K}=(X^0,\vec{X}).
\end{equation}
The Lorentz transformations for rotations around the $x$ and the $y$
axis are given by
\begin{eqnarray}
  \Lrotx{\phi} = 
  \left(
    \begin{array}{cccc}
      1 & 0 & 0 & 0 \\
      0 & 1 & 0 & 0 \\
      0 & 0 & \cos(\phi) & \sin(\phi) \\ 
      0 & 0 & -\sin(\phi) & \cos(\phi) \\ 
    \end{array}
  \right),\\ 
  \Lroty{\phi} = 
  \left(
    \begin{array}{cccc}
      1 & 0 & 0 & 0 \\
      0 & \cos(\phi)&0 & -\sin(\phi) \\ 
      0 & 0 & 1 & 0 \\
      0 &  \sin(\phi)&0 & \cos(\phi) \\ 
    \end{array}
  \right).
\end{eqnarray}
For the product we have
\begin{equation}
  \Lroty{\theta} \Lrotx{\phi} = 
  \left(\begin{array}{cccc}
    1&0&0&0 \\
    0&\cos{\theta}&\sin{\theta}\sin{\phi}&-\sin{\theta}\cos{\phi}\\
    0&0&\cos{\phi}&\sin{\phi} \\
    0&\sin{\theta}&-\cos{\theta}\sin{\phi}&\cos{\theta}\cos{\phi}\\
  \end{array}
\right).
\end{equation}

\bibliographystyle{JHEP}
\providecommand{\href}[2]{#2}\begingroup\raggedright\endgroup

\end{document}